\documentclass[journal]{IEEEtran}
\usepackage{enumitem}
\usepackage{graphicx} 
\usepackage{amsmath}
\usepackage{amssymb}
\usepackage{algorithm}
\usepackage{algpseudocode}
\usepackage{placeins}
\usepackage{subcaption}
\usepackage{lipsum}

\ifCLASSINFOpdf
\else
\fi
\begin{document}
%
\title{
Variational Source-Channel Coding for Semantic Communication
}
%
%
%
\author{Yulong Feng, Jing Xu, Liujun Hu, Guanghui Yu, and Xiangyang Duan

        
\thanks{Yulong Feng, Jing Xu, Liujun Hu, Guanghui Yu and Xiangyang Duan are with the State Key Laboratory of Mobile Network and Mobile Multimedia Technology, Shenzhen 518055, China, and ZTE Corporation, Nanshan District, Shenzhen 518055, China (e-mail: feng.yulong1@zte.com.cn, xu.jin7@zte.com.cn, hu.liujun@zte.com.cn, yu.guanghui@zte.com.cn, duan.xiangyang@zte.com.cn.).}
}

%
%

\markboth{Journal of \LaTeX\ Class Files,~Vol.~14, No.~8, August~2015}%
{Shell \MakeLowercase{\textit{et al.}}: Bare Demo of IEEEtran.cls for IEEE Journals}
%



\maketitle

\begin{abstract}
The rapid development of artificial intelligence (AI) brings new opportunities and challenges to classical communication. AI applications not only impose significant transmission pressure on current communication networks, but also affect the way information is stored and utilized because of its potent nonlinear fitting capabilities. Consequently, semantic communication technology emerges as a pivotal bridge connecting AI with classical communication. The current semantic communication systems are generally modeled as an Auto-Encoder (AE), while it lacks a deep integration of AI principles with communication strategies due to its inability to effectively capture channel dynamics. This gap makes it difficult to justify the need for joint source-channel coding (JSCC) and explain why performance improves. This paper begins by exploring lossless and lossy communication, highlighting that the inclusion of data distortion distinguishes semantic communication from classical communication. It breaks the conditions for the separation theorem to hold and explains why the amount of data transferred by semantic communication is less. Therefore, employing JSCC becomes imperative for achieving optimal semantic communication. Moreover, a Variational Source-Channel Coding (VSCC) method is proposed for constructing semantic communication systems based on data distortion theory, integrating variational inference and channel characteristics. Using a deep learning network, we develop a semantic communication system employing the VSCC method and demonstrate its capability for semantic transmission. We also establish semantic communication systems of equivalent complexity employing the AE method and the V‌ariational ‌A‌uto-‌E‌ncoder (VAE) method. The experimental results reveal that the VSCC model offers superior interpretability compared to the AE model, as it clearly captures the semantic features of the transmitted data, represented as the variance of latent variables in our experiments. In addition, the VSCC model exhibits superior semantic transmission capabilities compared to the VAE model. At the same level of data distortion evaluated by PSNR, the VSCC model exhibits stronger human interpretability, which can be partially assessed by SSIM.
\end{abstract}


%
\IEEEpeerreviewmaketitle

\section{Introduction}
\label{sec:intro} 

\IEEEPARstart{W}{ith} the continuous evolution of coding, modulation, waveform, and other communication technologies, classical communication has almost reached its theoretical bound~\cite{shannon1948mathematical}. Nevertheless, the emergence of numerous new applications, such as Extended Reality~\cite{zhang2023impact}, Immersive Communication~\cite{shen2023toward}, Large Language Models~\cite{zhao2023survey},~\cite{kasneci2023chatgpt}, etc., has made it difficult for classical communication networks to cope with the explosive growth of data transmission volumes~\cite{ji2021several}. These new applications, especially those based on Artificial Intelligence (AI), have also spurred the demand for intelligence in information transmission~\cite{hossain2023ai}. In the future, communication networks should actively integrate with AI to better and more seamlessly connect with various types of AI applications~\cite{wu2022ai}. It is necessary to study new communication methods and propose novel theories.

Current research in the field of communication appears to have moved beyond a narrow focus on traditional paradigms, increasingly depending on advancements in materials~\cite{aboagye2022ris}, devices~\cite{araniti2021toward}, and other technologies~\cite{liu2022survey}, while navigating the trade-offs between complexity, efficiency, and reliability. Much like 19th-century physics, the foundation of classical communication theory has been established, yet, not without ``clouds'' looming in the sky. The first challenge arises from the assumptions of classical information theory (CIT)~\cite{west2010introducing}, which impose significant limitations on the practical implementation of communication systems. In practice, human information often exhibits properties such as memory, non-stationarity, and non-ergodicity, which traditional statistical methods struggle to characterize effectively~\cite{cover1999elements}. For instance, human language is inherently ambiguous, encompassing subtleties like the ``tone'' of texts or the ``harmony'' of music~\cite{wang2005information}. The second challenge stems from the uncertainty introduced by AI. Since the powerful nonlinear data fitting capabilities of Artificial Neural Networks (ANNs) were demonstrated, particularly with recent large models excelling in natural language processing, it has become evident that problems previously considered intractable by statistical methods may be solved using ANNs~\cite{yarotsky2022universal}.

In fact, further advancements in communication theory were proposed shortly after the publication of Shannon's seminal paper. Weaver published another article on communication meaning~\cite{weaver1953recent}, in which he defined the concept of communication into three layers: \newline
\textbf{LEVEL A.} How accurately can the symbols of communica-tion be transmitted? (The technical problem.)\newline
\textbf{LEVEL B.} How precisely do the transmitted symbols convey the desired meaning? (The
semantic problem.)\newline
\textbf{LEVEL C.} How effectively does the received meaning affect conduct in the desired way?
(The effectiveness problem.)\newline
Weaver believed that CIT effectively addressed the LEVEL A problem. However, due to the highly mathematical abstraction, it overlooked two other crucial aspects. Resolving the LEVEL C problem depends on first addressing the LEVEL B problem. Thus, tackling the semantics problem is essential for breaking through the limitations of CIT. The most critical component in this endeavor is developing a semantic measurement method that is not constrained by the engineering assumptions in CIT. 

In 1953, Bar-Hillel and Carnap proposed the first systematic description and measurement of semantic information based on linguistic logic probability in~\cite{carnap1952outline}, focusing solely on propositions. It expanded the concept of information quantity beyond the limitations imposed by independently and identically distributed (i.i.d.) data, a framework commonly referred to as Classical Semantic Information Theory (CSIT). However, according to CSIT's formula, the semantic information of a contradictory proposition is infinite, which contradicts intuitive understanding. This is known as the BHC paradox~\cite{bar1953semantic}. Bar-Hillel and Carnap later explained the BHC paradox by arguing that contradictions contain infinite information and thus cannot be true. However, this explanation was considered unsatisfactory. Subsequent attempts like~\cite{bacchus1989representing} to describe information quantity in the vein of Bar-Hillel and Carnap failed to fully resolve this paradox. It wasn't until 2005 that Floridi recognized the fundamental flaw in CSIT~\cite{floridi2005semantic}. It fails to address the problem of semantic truthfulness. In CIT, this issue is naturally avoided by defining $0 \log 0 = 0$. CSIT also employs probability to describe semantic information, but when propositions are semantically untrue (e.g., contradictory propositions), the semantic information tends towards infinity. To address this issue, \cite{floridi2004outline} encapsulated semantic truthfulness within semantic information measurements, proposing a new framework known as the theory of Strongly Semantic Information (TSSI), which is based on truthfulness distance. However, the mathematical definition of TSSI is coarse, lacking the necessary deductive and logical structure between formulas, resulting in many combined propositions being unmeasurable. In 2011, \cite{d2011quantifying} proposed a semantic measurement theory that combines probability and distance, addressing the limitations of TSSI and extending its applicability. The most significant advancement in semantic communication came from the research by Bao et al.~\cite{bao2011towards}, published in 2011. They were the first to propose integrating local and common knowledge bases into semantic communication systems. They argued that by extending CSIT, one could provide a comprehensive description of semantic information. However, they also imposed certain assumptions regarding the applicability of semantic descriptions.

Building on these studies, it was recognized that traditional statistical theory-based methods might not fully capture semantic information. Researchers have begun to explore ANNs’ potential to integrate semantic information, with the continuous refinement of AI theory and the maturation of AI technologies. Theoretically, Kolchinsky et al.~\cite{kolchinsky2018semantic} defined semantic information as ``the syntactic information that a physical system has about its environment which is causally necessary for the system to maintain its own existence''. Kountouris et al.~\cite{kountouris2021semantics} envision a communication paradigm shift based on AI, which makes the semantics of information the foundation of the communication process. Their definition entails a goal-oriented unification of information generation, transmission, and reconstruction, by taking into account process dynamics, signal sparsity, data correlation, and semantic information attributes. Yuan et al.~\cite{yuan2022lossy} also employ AI to study semantic information. K\"{o}rner~entropy is employed to describe the number of AI states and semantic information~\cite{harangi2023generalizing}. Technically, numerous semantic communication systems have been developed based on various ANN models. The study initially appeared with joint source-channel coding (JSCC) based on deep learning, tracing back to Li et al.~\cite{rongwei2003joint}, in 2003. They attempted to replace source coding, channel coding, and modulation with Fully Connected Networks (FCNs). However, due to the limited development of AI technology at that time, the experimental results of the communication system were unsatisfactory. Subsequently, in 2019, Farsad et al.~\cite{farsad2018deep} incorporated a communication channel into the Seq2Seq model based on Long Short-Term Memory (LSTM) for JSCC in text transmission, achieving superior performance compared to traditional coding methods. The DeepJSCC system for image transmission was developed in 2020 by Deniz et al.~\cite{bourtsoulatze2019deep}, based on Convolutional Neural Networks (CNNs). This model, essentially a DCGAN, was trained using Auto-Encoder (AE) loss functions and exhibited performance comparable to classical communication while avoiding cliff effects. Based on these studies, models such as DeepJSCC-f~\cite{kurka2020deepjscc}, DeepJSCC-q~\cite{tung2022deepjscc}, DeepJSCC-1++~\cite{bian2023deepjscc} and JSCCformer-f~\cite{wu2024transformer} were developed to address feedback and modulation in practical communication. The concept of semantics was first introduced to explain JSCC gains in 2021 by Xie et al.~\cite{xie2021deep}. They built DeepSC model based on Transformer and FCN for text transmission, leading to the emergence of numerous similar models termed semantic communication systems. Semantic communication models have since been developed into various structures targeting different data types, such as the NTSCC (Nonlinear Transform Source Channel Coding) model for image transmission~\cite{dai2022nonlinear} and the SVC (Semantic Video Conferencing) model for video transmission~\cite{jiang2022wireless}.

\section{Problem Definition and Contributions}

\subsection{Problem Definition}

Nevertheless, semantic communication systems, modeled as AE trained end-to-end~\cite{wang2016auto}, lack the theoretical elucidation on the relationship with classical communication. The AE can be delineated by an encoding function $z=f_{\text{en}}(x)$ and a decoding function $\hat{x} = f_{\text{de}}(z)$. The encoder wants to compress data $x$ into a reduced-dimensional form $z$, which is called latent variable. The decoder can reconstruct $z$ back to its original state $x$, which echoes the objectives of classical communication. The loss function of AE typically encompasses only source and destination data, which make the training must be conducted end-to-end. Though gradients used to adjust encoder and decoder parameters are propagated back through the channel, which make these parameters intrinsically entwine with channel, the absence of explicit consideration for the channel hinders a clear interpretation of its role in the JSCC model~\cite{feng2023decoupling}. Consequently, semantic communication based on AE remains coarse, lacking modeling of the channel and source, failing to articulate the specific advantages of JSCC over traditional separate coding approaches. Regarding the current research on semantic communication implemented based on ANNs, fundamental questions persists and can be distilled as follows:
\begin{enumerate}[label=\arabic*]
    \item Where do the gains of semantic communication compared to classical communication come from?
    \item How does the implementation method of JSCC differ from separation coding in semantic communication?
    \item What role does the channel play in the JSCC framework of semantic communication?
\end{enumerate}

\subsection{Contributions}

In response to these three questions, this paper proposes a corresponding theoretical analysis. Based on the analysis, a new JSCC method and loss function are derived. An ANN model trained using this method, and the experimental results demonstrated its effectiveness. Consequently, this paper makes the following key contributions:
\begin{enumerate}[label=\arabic*]
\item By formally defining the core principle of tolerable data distortion, this work establishes a theoretical foundation for semantic communication, marking a fundamental departure from classical communication paradigms.

\item Through rigorous analysis grounded in CIT, the Lossy Coding Theorem, and the Information Bottleneck (IB) principle, we show that semantic communication must operate within the Joint Source-Channel Coding (JSCC) framework to attain optimal performance. Based on this insight, we further propose an semantic communication model.

\item A novel Variational Source-Channel Coding (VSCC) method is proposed that models the channel based on VAEs. In contrast to conventional VAEs, where the encoder deterministically generates latent representations, the proposed method incorporates the physical communication channel as an integral part of the encoding process via IB principle.

\item The VSCC model is implemented using ResNet Blocks and attention mechanisms, and is evaluated against AE-based and VAE-based semantic communication models with comparable parameter scales. The AE-based model serves as a representative of conventional semantic communication, while the VAE-based model acts as the algorithmic baseline for VSCC.

\item Experimental results show that semantic features remain consistent across varying transmitted data, yet these invariant features necessitate dynamic adaptation to fluctuating channel. It indicates that the latent representations in the VSCC model emerge  naturally from channel characteristics, rather than being artificially constructed by the encoder like VAE.

\item Experimental results show that the optimal Channel Matching Coefficient (CMC) in the loss function varies with the channel SNRs, suggesting that the channel functions as a part of the joint encoder and contributes to the overall data distortion. The proposed VSCC framework balances the trade-off between semantic distortion and channel noise by dynamically adapting to both the source characteristics and channel conditions.

\item Experimental results show that the proposed VSCC model consistently outperforms VAE-based approaches in terms of both Peak Signal-to-Noise Ratio (PSNR), which primarily reflects reconstruction quality, and Structural Similarity Index (SSIM), which serves as a proxy for semantic fidelity. Moreover, under challenging channel conditions, VSCC retains performance advantages over conventional AE-based semantic communication models.
\end{enumerate}

The remainder of this paper is organized as follows. In the next section, the aforementioned three issues will be addressed. A detailed explanation of the relationship between semantic communication and classical communication is provided, demonstrating the rationale for implementing semantic communication based on the JSCC. Building upon this research, a VSCC method matched to the channel is proposed. In Section IV, the VSCC model implemented using ANNs is elaborately introduced, along with the associated training and testing algorithms. Section V offers a comprehensive overview of the experimental results based on the proposed VSCC method, in comparison with the AE method and the VAE method. And a careful analysis of the experimental results is conducted, providing explanations and demonstrating the validity of the semantic communication definitions. In Section VI, the entire paper is summarized, while also highlighting potential shortcomings in the current experiments and identifying directions for future research.

\section{Theoretical Analysis}

This section explores the construction of semantic communication and its relationship with classical communication. It provides the rationale for adopting the JSCC in semantic communication and explains how to model the channel within JSCC using variational inference.

\subsection{Lossless and Lossy Transmission}

To address the first research question outlined in Section~\ref{sec:intro}, this paper builds upon existing semantic measurement studies to propose a fundamental assumption that data information and semantic information are not one-to-one correspondences. Data serves as the carrier for semantic information, but not the entirety of data information is necessary to convey semantics. For instance, consider the sentences ``Bai is not a cat, it's a dog'' and ``Bai is a dog'', where the former contains more data information than the latter, yet both convey the same semantics. Therefore, in semantic communication, data distortion at the source is necessary~\cite{shannon1959coding}. This contradicts the source coding theorem in CIT, which aims for lossless communication. Semantic communication cannot be effectively evaluated with bit error rate~\cite{gunduz2022beyond}. It is based on rate-distortion theorem, and exhibits greater tolerance for data errors, which explains its main differences from classical communication and the gain source. Consequently, we define semantic communication as allowing data distortion~\cite{sims2016rate},~\cite{berger2003rate}, yet achieving semantic fidelity through the establishment of prior information (knowledge base). And this assertion highlights the need to distinguish between lossless communication and lossy communication.

\begin{figure*}
    \centering
    \includegraphics[width=1\linewidth]{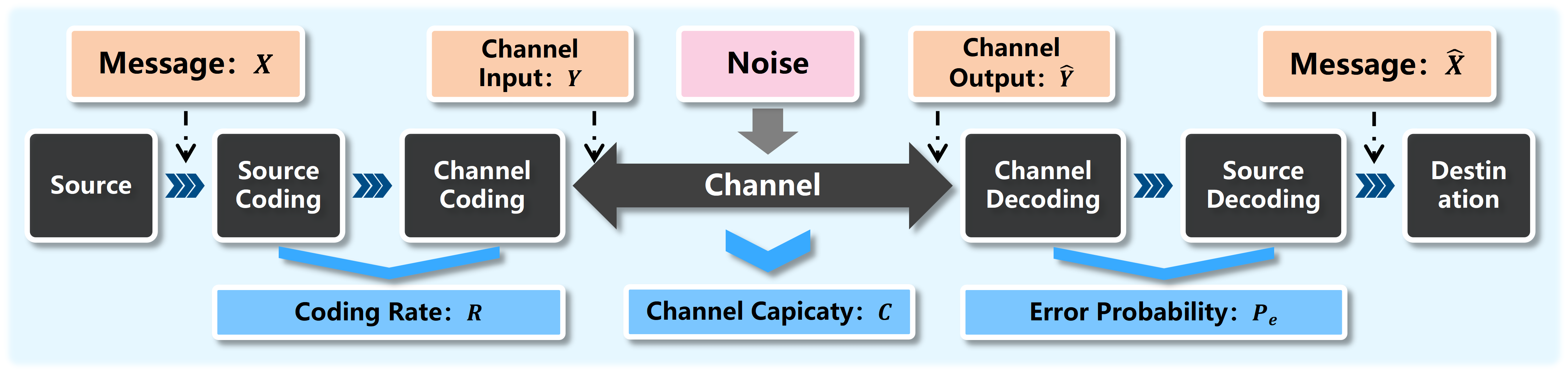}
    \caption{Communication model based on classical communication theory. Basically, it consists of a source module, an encoding module, a channel module, a decoding module, and a destination module. The encoding module is divided into source encoding and channel encoding, which are guaranteed by Shannon's source coding theorem, Shannon's channel coding theorem, and the source-channel separation theorem.}
    \label{fig:1}
\end{figure*}

As shown in Fig.~\ref{fig:1}, let $X$ denote the source, $P(X)$ denote the distribution of the source, $H(X)$ denote the entropy of the source, and $R$ denote the rate of coding. Assuming that the source symbols belong to the typical set and disregarding their specific meaning, Shannon's source coding theorem can be derived.

\textit{\textbf{Theorem 1 (Shannon's Source Coding Theorem ):} For a given source $X$ and encoding rate $R$, if $R>H(X)$, $R$ is accessible; If $R<H(X)$, $R$ is inaccessible.}

A discrete memoryless channel (DMC) $\{Y,p(\hat{y}|y),\hat{Y}\}$ has a capacity denoted by $C$, where $Y$ represents the channel input and $\hat{Y}$ represents the channel output. The samples of $Y$ and $\hat{Y}$ are $y$ and $\hat{y}$, respectively. The conditional probability $p(\hat{y}|y)$ describes the channel transition probability. When $y$ and $\hat{y}$ form a jointly typical set, there exists Shannon's channel coding theorem.

\textit{\textbf{Theorem 2 (Shannon's Channel Coding Theorem ):} For any coding rate $R<C$, there exists a sequence of codewords $(2^{nR},n)$, when $n \to \infty$, achieving the probability of error $P_{e}^{(n)} \to 0$. On the contrary, if the bit error rate $P_{e}^{(n)} \to 0$ of the codewords $(2^{nR},n)$, there must be $R<C$.}

It is worth noting that the aforementioned source coding theorem is independent of the channel noise distribution, and the channel coding theorem is independent of the source data distribution. Therefore, according to Shannon's source-channel coding theorem, the conclusion is reached that if the original source is transmitted without distortion, separated source-channel coding (SSCC) is equivalent to JSCC, and both can achieve optimal transmission.

\textit{\textbf{Theorem 3 (The Source-Channel Coding Theorem):} If $x_1,x_2,...x_n$ is a stochastic process satisfying asymptotic equipartition property (AEP) and belongs to a finite alphabet $X$ with $H(X)<C$. There is a source-channel coding method make the error probability $P_{e}^{(n)} \to 0$. Conversely, for any stationary stochastic process, if $H(X)>C$, then the error probability is far from 0, and it is impossible to send this source over the channel with an arbitrarily low error probability.}

However, these theorems are designed for lossless data transmission. When distortion is introduced to the original source, the communication process can no longer be adequately described by these three theorems alone. For semantic communication, semantics invariably depend on the existence of data, but data does not necessarily convey semantics. When the semantic encoder extracts semantic features from data, it removes data redundancies, making semantic communication potentially more efficient than classical communication methods. Building on this premise, we argues that semantic communication functions within the framework of rate distortion theory for data, achieving semantic fidelity in information transmission through the effective utilization of the knowledge base.

\textit{\textbf{Theorem 4 (Shannon’s Lossy Source Coding Theorem):} For an independent and identically distributed (i.i.d.) information source $X$ with a distribution of $P(X)$, if the decoded source is represented as $\hat{X}$ and the distortion function is represented as $d(x,\hat{x})$, then the rate-distortion function is:}
\begin{equation}
R(D)=\min _{P(\hat{x} \mid x) \in P_D} I(X ; \hat{X}).
\end{equation}
It represents the minimum information rate that can be achieved under average distortion $D$. Where $D = E[d(m,\hat{m})]$ represents the average distortion, $P_D$ represents any probability distribution that satisfies the average distortion $D$, and $I(M;\hat{M})$ represents the mutual information between the transmitted and received messages.

\FloatBarrier
\begin{figure*}
    \centering
    \includegraphics[width=1\linewidth]{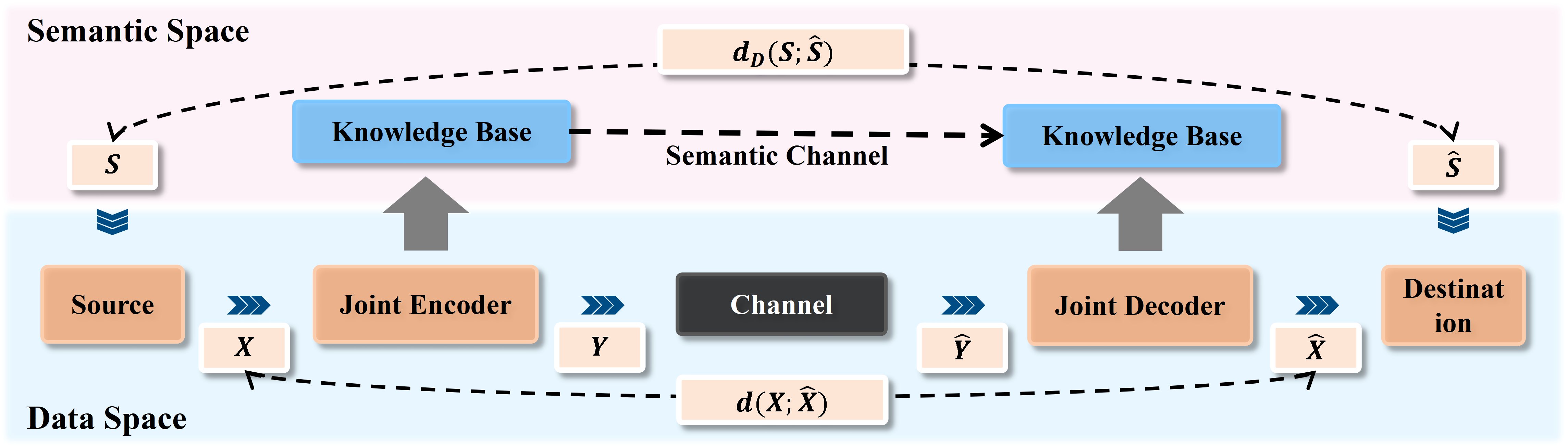}
    \caption{Semantic communication model. It is achieved through theoretical analysis in the semantic space and implementation in the data space. The key distinction between the two spaces lies in whether the causality of the physical world is considered. The data space communication model is similar to the classical communication model but utilizes a joint encoding module and a decoding module. The semantic space is primarily defined by a knowledge base (KB), which governs how data is mapped to the semantic space and aids in the encoding and decoding process. The channel that aligns the KBs can be viewed as the semantic channel.}
    \label{fig:2}
\end{figure*}

In practical communication processes, source encoding is inherently lossy. This is due to the impracticality of accurately calculating the actual data entropy, especially for images. The challenge arises because the data being transmitted typically cannot effectively represent the entire population, and thus the estimated distribution does not accurately reflect the population distribution. Moreover, the goal of practical communication is generally to achieve better transmission performance in typical communication scenarios rather than to optimize over an infinitely long time. As a result, data is not considered from the perspective of a typical set, and the source does not exhibit the asymptotic equipartition property (AEP). The effectiveness of data compression and recovery is not simply measured by the bit error rate, but rather by evaluating the source encoding and decoding in terms of the content of the data within the current communication process. Therefore, even without considering channel matching, semantic has already been ``applied'' in practical communication processes.

\begin{figure*}
    \centering
    \includegraphics[width=0.9\linewidth]{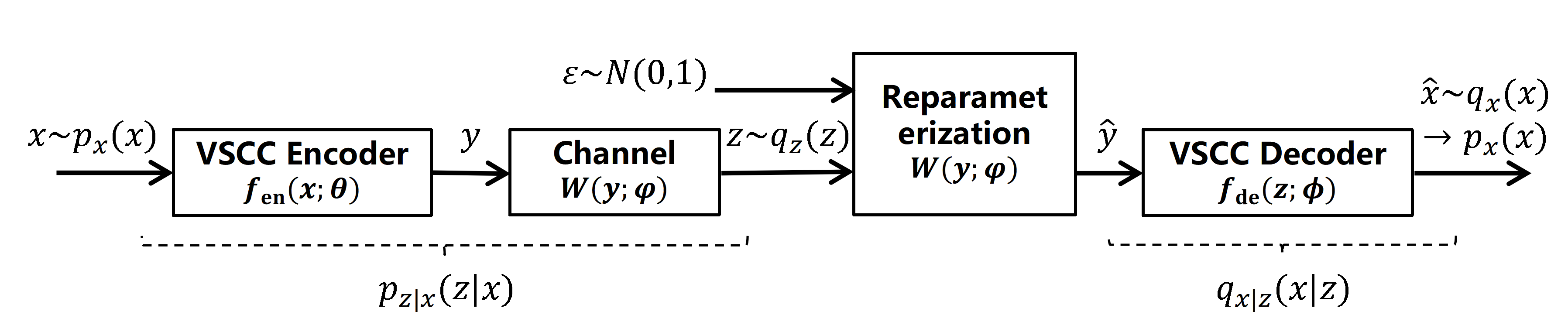}
    \caption{Diagram illustrating the mathematical reasoning of the VSCC method. The purpose of this method is to make the received data distribution \( q_x(x) \) as same as the original distribution \( p_x(x) \) as possible. The original data $x$ is first passed through the VSCC encoder to obtain the encoding vector $y$. The edcoded vector $y$ passes through the channel to obtain the hidden variable $z$. $z$ stands for feature distributions that can be combined into the original data distribution. The receiver resamples $\hat{y}$ from the hidden variable $z$ and then decodes it to the data $\hat{x}$.}
    \label{fig:3}
\end{figure*}

\subsection{Joint and Separated Source-Channel Coding}

Building on the investigation of the first question in section~\ref{sec:intro}, we further explores the second question. While the lossy transmission theory is not yet fully developed, existing proofs indicate that JSCC can still achieve optimality~\cite{gastpar2003code}. JSCC considers not only the source and channel noise distributions but also the relationship between source distortion and channel noise. In most cases of source distortion, the source encoding rate and channel encoding rate in SSCC lack closed-form analytical expressions, making it impossible to achieve optimality in practice~\cite{gastpar2002code}. Therefore, we propose a semantic communication model based on the JSCC framework that jointly optimizes knowledge base integration and effectiveness problem. This approach can provide advantages over SSCC when striving for optimal transmission under practical constraints.

When data compression is described by the rate-distortion function instead of Shannon's source coding theorem, the independence between the source and channel determines whether JSCC has an advantage over SSCC. In fact, Shannon also investigated this question when he proposed the rate-distortion theorem. He found that JSCC can still achieve optimality. However, he did not further analyze the specific conditions under which JSCC achieves optimality or whether SSCC remains equivalent to it.

The data transmission described by the rate-distortion function no longer focuses solely on matching the source distribution \( P(X) \) with the channel transition distribution \( P(\hat{Y}|Y) \), as in CIT. Instead, it integrates the distortion function \( d(x,\hat{x}) \), source distribution \( P(X) \), and channel transition distribution \( P(\hat{Y}|Y) \). Consequently, optimal coding must not only minimize the distortion rate but also consider whether the channel noise can be leveraged as part of the distortion function. Subsequent studies~\cite{gastpar2003code} have elaborated on this approach.

\textit{\textbf{Theorem 5:} For a given source distribution $P(X)$, channel condition distribution $P(\hat{X}|X)$, and corresponding single-letter encoding method $(f,j)$, when $I(X;\hat{X})>0$, if and only if the distortion function satisfies:
\begin{equation}
d(x ; \hat{x})=-c \log _2 p(x \mid \hat{x})+d_0(x),
\label{eq:2}
\end{equation}
where $c>0$, and $d_0$ is the function of $x$, the communication system can achieve the optimal transmission.}

Theorem 5 suggests that communication systems described by Theorems 3 and 4 can achieve optimal performance with JSCC, provided that the distortion function satisfies the Eq.\ref{eq:2}. The Eq.\ref{eq:2} reflects the source distribution and the channel characteristics, indicating that the JSCC method accounts for the coordination between source distortion and channel noise. In semantic communication, the joint encoder facilitates this coordination by transforming the original source distribution into one that aligns with the channel transition distribution. Consequently, in scenarios involving data distortion, semantic communication may offer additional advantages over classical communication in the context of JSCC.

However, is JSCC truly indispensable for semantic communication? Unlike CIT, which benefits from the separation theorem, semantic communication lacks a well-established mathematical framework to support such endeavors. Recent studies~\cite{gastpar2002code} have shown that finding an analytical solution for the source distortion rate that perfectly matches a fixed channel coding rate, or vice versa, is unattainable. Nevertheless, if based on the JSCC framework, the IB theory~\cite{tishby2000information} might provide the best interpretation for semantic communication systems. The IB describes the optimal representation of information during data compression and feature extraction. Its core idea is to retain the information most relevant to the target task (i.e., transmitting semantic information) while compressing redundant information from the raw data (i.e., reducing data transmission). Given input data $ X $ and target task $ \hat{X} $, we seek an intermediate representation $ \hat{Y} $ that maximizes its predictive capability for $ \hat{X} $ while minimizing its dependence on $ X $. This can be achieved by minimizing Eq.~\ref{eq:ib}:

\begin{equation}
\mathcal{L}_{\text{IB}}[p(z|x)] = I(X;\hat{Y}) - \beta I(\hat{Y};\hat{X}).
\label{eq:ib}
\end{equation}

The IB problem actually corresponds to the semantic communication model as depicted in Fig.\ref{fig:2}. It is the proposed semantic communication model based on JSCC. While semantic information $S$ is transmitted in the data space, its theoretical measurement occurs in the semantic space. The data transmission model resembles the classical communication model, but with SSCC replaced by JSCC, potentially providing a better alignment between source data distortion $d(X;\hat{X})$ and channel noise. Additionally, semantic communication involves the deployment of a knowledge base and the consideration of communication objectives within the semantic space. The knowledge base, serving as a repository of prior information, bolsters the semantic encoding and decoding processes. Furthermore, for diverse communicative tasks, semantic information may also introduce distortions $d_{D}(S;\hat{S})$, retaining only the information relevant to the task, which suggests that semantic communication has the potential to further reduce data transmission. Consequently, the realization of semantic communication means we aim to ensure that the received information contains as much of the original semantic meaning as possible, which corresponds to the process of maximizing $\beta I(\hat{Y};\hat{X})$ in Eq.\ref{eq:ib}. While achieving optimal semantic communication requires a delicate balance among various factors, including data distortions, source distribution, channel transition distribution, the knowledge base, and semantic distortions. It means we want the received information to be as compact as possible, which corresponds to the process of minimizing $I(X;\hat{Y})$ in Eq.\ref{eq:ib}.

Traditional statistical theories lack the necessary tools to analyze such complex relationships, leveraging JSCC based on ANNs may offer a viable approximation. In deep learning, the Variational Information Bottleneck (VIB) ~\cite{alemi2016deep} is typically employed as a loss function to train ANNs and achieve IB-like results. The VIB loss function is given by:

\begin{equation}
\begin{aligned}
\mathcal{L}_{\text{VIB}} = 
& ~ \mathbb{E}_{x\sim p(x)} \mathbb{E}_{\hat{y}\sim p(\hat{y}|x)} \left[ -\log q(\hat{x}|\hat{y}) \right]
\\
& + \beta \cdot 
\mathbb{E}_{x\sim p(x)} \left[ \text{KL}\big( p(\hat{y}|x) \parallel q(\hat{y}) \big) \right].
\label{eq:vib}
\end{aligned}
\end{equation}
Notably, if we set $ q(\hat{y}) $ to be a standard normal distribution, and replace $\hat{y}$ with $z$, that is the standard latent variable symbol in VAEs, Eq.~\ref{eq:vib} becomes nearly identical to the loss function of a VAE, with the addition of a hyperparameter $ \beta $. In the next section, we present our VSCC model derived through variational inference rather than from the IB perspective. This approach is taken because the IB theory primarily provides conceptual guidance rather than specific methodologies for modeling ANN loss functions. Variational inference offers a more effective way to incorporate channel modeling into the encoder and explain its role in communication.

\subsection{Variational Source-Channel Coding for Channel Matching}

The exploration of the two preceding questions elucidates that effective semantic communication requires explicit channel modeling in the JSCC. Since semantic communication inherently involves data distortion to eliminate redundant information unnecessary for semantic expression, lossy encoding is applied at the source. According to the optimal theorem for JSCC under lossy conditions, the lossy source in semantic communication could be achieved through the channel noise. Consequently, the channel no longer acts as an obstacle that necessitates additional data redundancy, such as channel encoding, to offset its impact, but instead becomes an integral component of the semantic coding process. To effectively utilize channel for distorting source data, appropriate matching between the channel noise distribution and the source data distribution is necessary. This alignment essentially transforms the original source data distribution into a new distribution, facilitating controlled distortion of data through the noise distribution, while simultaneously preserving the essential characteristics of the original data distribution. Nevertheless, the new distribution may lack a closed-form analytical expression, because data distribution transformation has always been a challenge in statistics~\cite{stuart2010kendall}. Fortunately, research on ANNs offers effective approximation methods for addressing these challenges~\cite{dinh2014nice},~\cite{ho2020denoising}. Building on variational inference and VIB, this paper proposes a VSCC method that integrates channel state information into the training function of semantic communication, thereby achieving the necessary alignment process.

\begin{figure*}
    \centering
    \includegraphics[width=0.8\linewidth]{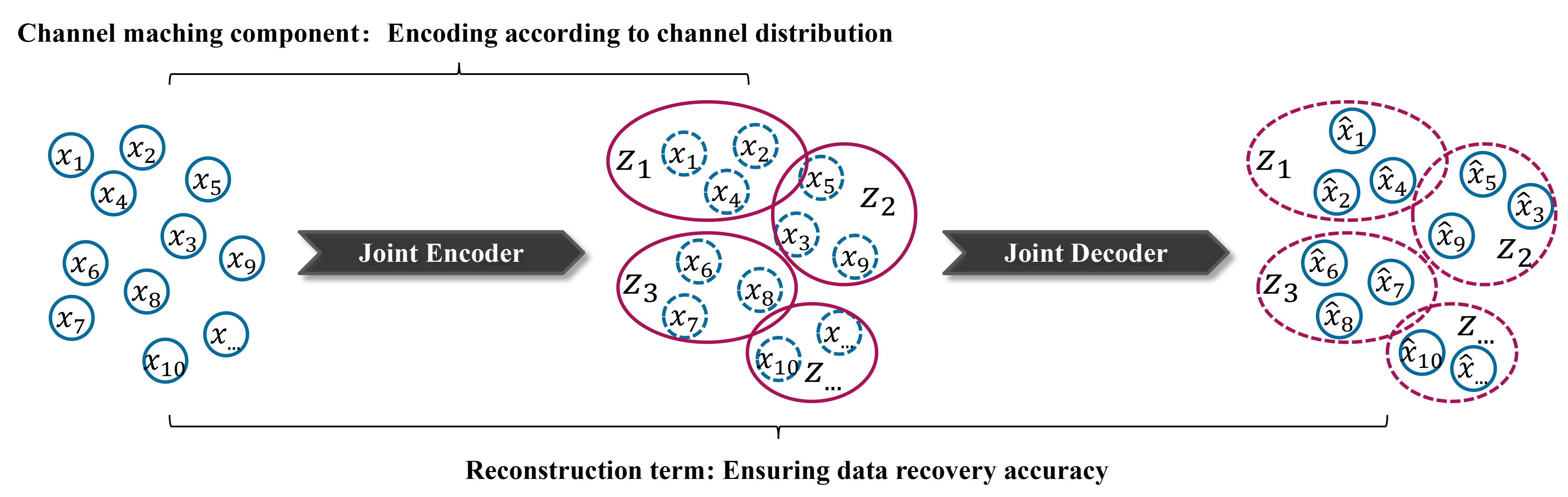}
    \caption{Geometric interpretation of the Variational Source-Channel Coding (VSCC) method. The raw data \(x\) is a single sample from the original data distribution. It can be encoded by VSCC encoder and the channel into the latent variable \(z\). Each feature distribution represented by \(z\) corresponds to a cluster of \(x\), forming distinct regions. By resampling different feature distributions, the original data distribution can be reconstructed and represented by \( \hat{x} \).}
    \label{fig:4}
\end{figure*}

By integrating variational inference and VIB principle with communication theory, the VSCC method establishes a robust algorithmic foundation for training semantic communication systems. As shown in Fig.~\ref{fig:3}, the VSCC method aims to recover the original source data distribution, which is denoted as $p_x(x)$. The received data distribution is represented as $q_x(x)$. This objective is achieved through two steps. 

In the first step, following the principle of maximum likelihood, the goal is to minimize the negative logarithm of the likelihood between these two distributions:
\begin{equation}
\min \mathbb{E}_{x \sim p_x(x)}[-\log q_x(x)].
\label{eq:3} 
\end{equation}

Given that $p_x(x)$ is known, minimizing Eq.~\ref{eq:3} is equivalent to minimizing the Kullback-Leibler (KL) divergence between $p_x(x)$ and $q_x(x)$.
\begin{equation}
\begin{aligned}
& \text{KL}\left(p_x(x) \| q_x(x)\right) \\
= \ & \mathbb{E}_{x \sim p_x(x)} \left[\log p_x(x)\right] - \mathbb{E}_{x \sim p_x(x)} \left[\log q_x(x)\right] \\
= \ & C - \mathbb{E}_{x \sim p_x(x)} \left[\log q_x(x)\right],
\end{aligned}
\label{eq:4}
\end{equation}
where $C$ is a constant.

To ensure coherent alignment between the channel and the source, a key step in the VSCC method involves treating the output variables from an intermediate channel layer as latent variables $z$. These latent variables are used to facilitate a matching process specifically tailored to the characteristics of the channel, thereby enhancing the interpretability of JSCC. Thus, the KL divergence of the joint probability density is introduced with the inclusion of the latent variable $z$, and an upper bound can be derived as follows:
\begin{equation}
\begin{aligned}
& \text{KL}\left(p_{x, z}(x, z) \| q_{x, z}(x, z)\right) \\
= \ & \text{KL}\left(p_x(x) \| q_x(x)\right) \\
& \quad + \int_{x} p_x(x) \, \text{KL}\left(p_{z \mid x}(z \mid x) \| q_{z \mid x}(z \mid x)\right) \, dx \\
\geq \ & \text{KL}\left(p_x(x) \| q_x(x)\right).
\end{aligned}
\label{eq:5}
\end{equation}

Therefore, minimizing Eq.~\ref{eq:5} is equivalent to minimizing the upper bound of Eq.~\ref{eq:3}, implying that maximum likelihood estimation can be achieved by minimizing Eq.~\ref{eq:5}. The expansion of Eq.~\ref{eq:5} is as follows:
\begin{equation}
\begin{aligned}
& \min \text{KL}\left(p_{x, z}(x, z) \| q_{x, z}(x, z)\right) 
\\
= \ & \min \iint p_x(x) \, p_{z \mid x}(z \mid x) \log \frac{p_x(x) \, p_{z \mid x}(z \mid x)}{q_z(z) \, q_{x \mid z}(x \mid z)} \, dx \, dz 
\\
= \ & \min \mathbb{E}_{x \sim p_x(x)} \left[ \int_{z} p_{z \mid x}(z \mid x) \log \frac{p_{z \mid x}(z \mid x)}{q_z(z) \, q_{x \mid z}(x \mid z)} \, dz \right] + C,
\end{aligned}
\label{eq:6}
\end{equation}
where $C$ is a constant.

Thus, Eq.~\ref{eq:6} can be transformed to:
\begin{equation}
\begin{aligned}
& \min \text{KL}\left(p_{x, z}(x, z) \| q_{x, z}(x, z)\right) 
\\
 \Leftrightarrow \ & \min \mathbb{E}_{x \sim p_x(x)} \left[ \int_{z} p_{z \mid x}(z \mid x) \log \frac{p_{z \mid x}(z \mid x)}{q_z(z) \, q_{x \mid z}(x \mid z)} \, dz \right] 
\\
= \ & \min \mathbb{E}_{x \sim p_x(x)} \big\{ \text{KL}\left(p_{z \mid x}(z \mid x) \| q_z(z)\right) 
\\
& \quad + \mathbb{E}_{z \sim p_{z \mid x}(z \mid x)} \left[ -\log q_{x \mid z}(x \mid z) \right] \big\}.
\end{aligned}
\label{eq:7}
\end{equation}
It is worth noting that the Eq.~\ref{eq:7} is nearly identical to the VIB Eq.~\ref{eq:vib}, with the only difference being the weighting factor $\beta$. This confirms the equivalence between variational inference and the VIB principle. The weight $\beta$ controls the degree of information compression in the VIB, which will be reflected in the second step through the CMC.

In the second step, suppose the output from the joint encoder, denoted as $y$, follows a Gaussian distribution $\mathcal{N}(\mu_1,\sigma_1^2)$, and the channel layer introduces Additive White Gaussian Noise (AWGN) with distribution $\mathcal{N}(0,\sigma_2^2)$. The channel output is given by:
\begin{equation}
z = W\left(f_{\text{en}}(x; \theta); \varphi\right) = f_{\text{en}}(x; \theta) + n,
\label{eq:8}
\end{equation}
where $W$ represents the channel function parameterized by $\varphi$, while $f_{\text{en}}$ denotes the encoder function with parameter $\theta$. It can be easily found $\varphi$ only includes $\sigma_2$ for AWGN.

It follows that $p_{z|x}(z|x) = \mathcal{N}(\mu=\mu_1, \sigma^2=\sigma_1^2+\sigma_2^2)$. Additionally, we assume that the true latent variable $z$ follows a distribution similar to the channel noise. Specifically, let $q_{z}(z) = \mathcal{N}(0, \sigma_2^2 + d)$, where $d$ is the CMC. The CMC is assumed to represent the true variance of the encoded vector $y$. At the same time, the CMC also reflects the role of the parameter $\beta$ in Eq.~\ref{eq:vib} of the VIB in controlling the degree of information compression, with the distinction that $\beta$ operates multiplicatively, whereas the CMC functions additively.

In order to generate source coding vectors according to channel SNR, CMC should be adjusted based on the SNR of the channel. Substituting $p_{z|x}(z|x)$ and $q_{z}(z)$ into Eq.~\ref{eq:7} yields:
\begin{equation}
\begin{aligned}
& \min \text{KL}\left(\mathcal{N}\left(\mu_1, \sigma_1^2 + \sigma_2^2\right) \| \mathcal{N}\left(0, \sigma_2^2 + d \right)\right) 
\\
= \ & \min \frac{1}{2} \left( \log \frac{\sigma_2^2 + d}{\sigma_1^2 + \sigma_2^2} + \frac{\mu_1^2 + \sigma_1^2 + \sigma_2^2}{\sigma_2^2 + d} - 1 \right).
\\
\end{aligned}
\label{eq:9}
\end{equation}

Thus, we can get the loss function of the VSCC:
\begin{equation}
\begin{aligned}
& \mathcal{L}_{\text{VSCC}} = \mathbb{E}_{x \sim p(x)} \bigg[ \frac{1}{2} \big( \log \frac{\sigma_2^2 + d}{\sigma_1^2(x;\theta) + \sigma_2^2}
\\
& + \frac{\mu_1^2(x;\theta) + \sigma_1^2(x;\theta) + \sigma_2^2}{\sigma_2^2 + d} - 1 \big) - \log q_{x \mid z}(x \mid z) \bigg]. 
\end{aligned}
\label{eq:10}
\end{equation}
The first term in Eq.~\ref{eq:10} serves as the channel matching component, facilitating the alignment of the encoded vector with the channel transition distribution. The second term, referred to as the reconstruction term, ensures that the received data distribution closely matches the original data distribution.

On one hand, unlike VAE, the VSCC method uses the first term in Eq.~\ref{eq:10} to align the latent variable $z$ as closely as possible with the channel noise distribution, rather than a standard Gaussian distribution. This alignment implies that the latent variables $z$ are derived from the original data $x$ by superimposing channel noise. As a result, the channel is no longer an obstacle to be overcome but becomes an integral part of the encoding process. The channel noise aids in eliminating redundancy in $x$, leading to different feature distributions represented by the latent variables $z$. If these feature distributions are considered to be the semantic information of the original message distribution, then the VSCC method effectively achieves both the elimination of data redundancy and the extraction of semantic information.

On the other hand, unlike AE, the VSCC method does not require the original data $x$ to be identical to the received data $\hat{x}$. Instead, it utilizes the second term of Eq.~\ref{eq:10} to ensure that the original data distribution and the received data distribution are as similar as possible. This approach not only constrains the encoder to encode based on the distributional features of the original message, resulting in specific feature distributions, but also guides the decoder to reconstruct data that aligns with the distributional features of the original message. This dual consideration ensures that both the encoder and decoder are optimized for semantic fidelity, within the constraints imposed by the channel.

It is noteworthy that the reconstruction term in Eq.~\ref{eq:10} has been simplified by sampling the latent variable only once. This simplification is adopted under the premise of relatively straightforward experimental settings and assumptions. However, in scenarios involving more complex channel conditions, multiple samplings of the latent variable might be necessitated.

\begin{figure*}
    \centering
    \includegraphics[width=1\linewidth]{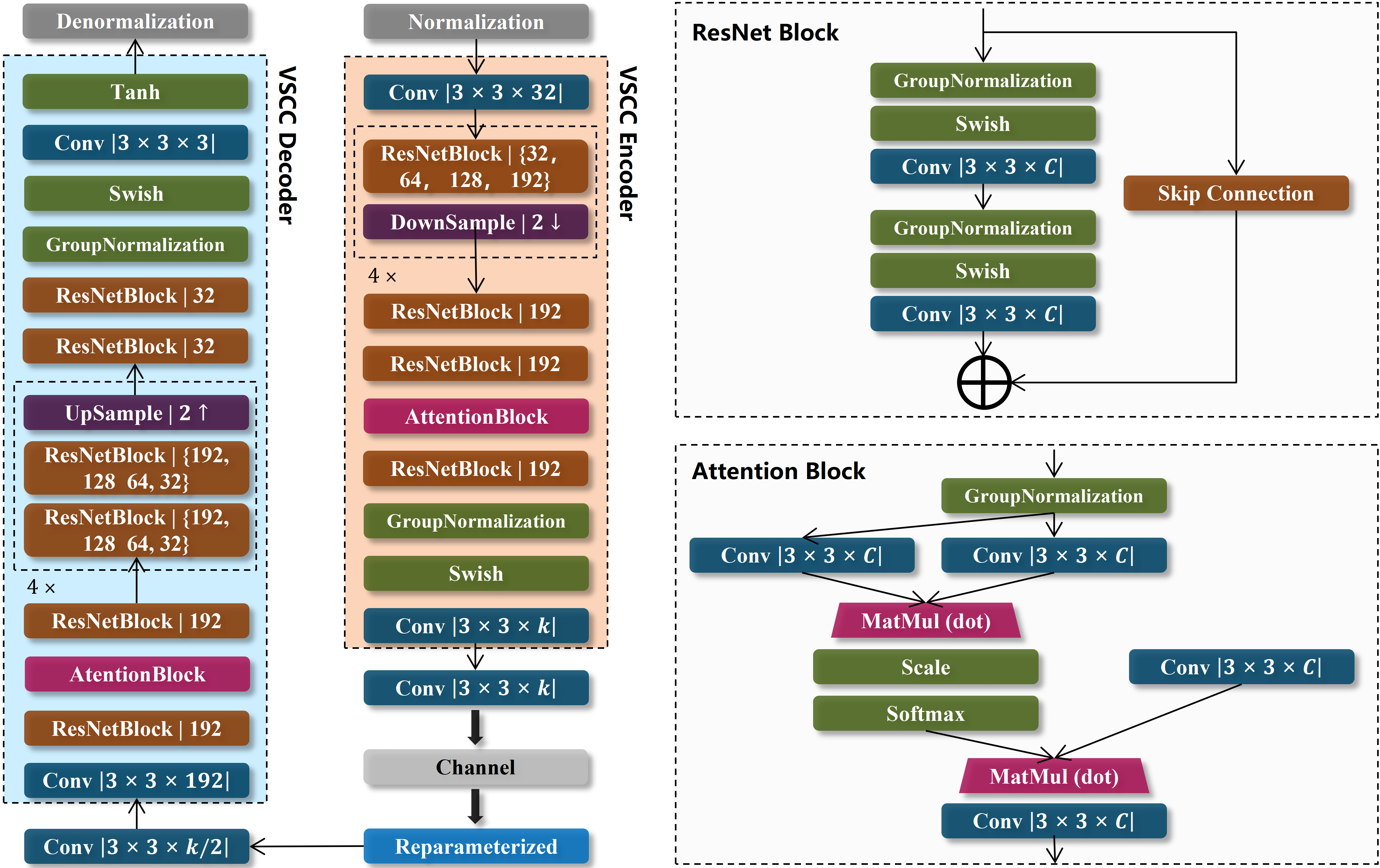}
    \caption{The structure of the semantic communication model, constructed using Residual Blocks and Attention Blocks. The model can be trained using Auto-Encoder (AE), Variational AE (VAE), or Variational Source-Channel Coding (VSCC) methods. In the joint encoder, the image \(X\) is first normalized and processed through a Convolutional Neural Network (CNN) layer to increase the feature dimension to 32, followed by a sequence of four Residual Blocks and Attention Blocks to extract features across various dimensions. Additional Residual and Attention Blocks finalize the feature dimensions before the data is compressed with a Group Normal (GN) layer, Swish activation function, and two CNN layers. The feature dimension \( k \) of encoded vector \( Y \) is set as 16. In the decoder, the received latent variable \(Z\) is resampled using a reparameterization module and decoded into \(\hat{X}\) by reversing the steps of the joint encoder. The Residual Block structure (top right) includes a GN layer, Swish activation function, and a CNN layer to enhance input features, while the Attention Block structure (bottom right) computes the output using the attention mechanism with inputs \(Q\), \(K\), and \(V\) produced by CNN layers.}
    \label{fig:5}
\end{figure*}

Consequently, refer to Fig.~\ref{fig:3}, the original message $X = \{x_1, x_2, \dots\}$ is first processed by the joint encoder to produce $Y$. This encoded vector then passes through the channel, resulting in the latent variable $Z = \{z_1, z_2, \dots\}$. At the receiver, samples $\hat{Y}$ are drawn from $Z$ and decoded to obtain $\hat{X} = \{\hat{x}_1, \hat{x}_2, \dots\}$. The VSCC method treats the original data $X$ as a fixed instance, representing a sample from the original message distribution. Latent variable $Z$ is a random variable associated with a specific feature distribution. The encoding process decomposes the original message distribution into distinct feature distributions, similar to representing any arbitrary distribution as a combination of multiple Gaussian distributions.

Geometrically, as shown in Fig.~\ref{fig:4}, each data point $x$ is a fixed sample from the original message distribution. Each latent variable $z$ is a random variable representing a small region containing multiple samples $x$ with the same distribution. In this manner, the VSCC encoder classifies the original message $x$ based on different feature distributions. The receiver then resamples from the distribution represented by $z$ and uses the resampled data $\hat{y}$ to reconstruct $\hat{x}$. Although the reconstructed data $\hat{x}$ shares the same distribution as the original data $x$, the individual data points may differ.

\section{Architecture and Implementations}

To demonstrate the VSCC method's capability to align with the channel, we develop an end-to-end semantic communication model based on ResNet blocks, as illustrated in Fig~\ref{fig:5}. The ResNet blocks, with their varying receptive fields, extract multi-dimensional features from the image. These features are encoded as latent variables. By resampling different feature distributions represented by latent variables, the alignment between the source and the channel can be achieved.

\subsection{Semantic Communication Model}

The semantic communication model consists of a joint encoder and a joint decoder, with the channel acting as an integral part of the joint encoder in the VSCC method. The data transmission process is described as follows:

The image to be transmitted, denoted as $X \subset \mathbb{R}^{H \times W \times C}$, has pixel values ranging from 0 to 255. Since DNNs typically process data within the [0, 1] range to align different data features and accelerate convergence, $X = \{x_i; i = 1, 2, \dots, H \times W \times C\}$ is first normalized. The normalized image then enters the joint encoder $f_\text{en}(x; \theta)$, where $\theta$ represents the trainable parameters. The joint encoder initially uses a CNN layer to increase the feature dimension $C$ of the image to 32. Subsequently, the image passes through four ResNet blocks and DownSample layers. In each cycle, the ResNet block increases the feature dimension $C$ to 32, 64, 128, and 192, while the DownSample layers halve the height $H$ and width $W$ of the image. The DownSample layers are implemented using CNNs, compressing the image size and extracting its features. Each cycle processes images with different heights $H$, widths $W$, and feature dimensions $C$, enabling feature extraction across various dimensions. Following this, three additional ResNet blocks and an Attention block capture the remaining feature information in the 192-dimensional space. Finally, a CNN layer compresses the normalized feature information into a encoded vector $Y$, which can be regarded as following the distribution $\mathcal{N}(\mu_1, \sigma_1^2)$. This encoded vector are then transmitted through the channel.

Before transmission through the channel, a CNN layer transforms the encoded vector $Y$ to better match the channel characteristics. Assuming the channel noise follows a Gaussian distribution $\mathcal{N}(0, \sigma_2^2)$, the received latent variable $Z$ will follow a distribution $\mathcal{N}(\mu_1, \sigma_1^2 + \sigma_2^2)$. Before decoding, $Z$ passes through a reparameterization layer. This layer, which contains no learnable parameters, resamples a new latent variable $\hat{Y}$ based on $Z$, ensuring that $\hat{Y}$ captures both channel and source characteristics. Finally, $\hat{Y}$ is mapped into a latent space suitable for decoding by a CNN layer.

The latent variable $\hat{Y}$ is decoded by the joint decoder $f_\text{de}(z;\phi)$, parameterized by $\phi$. The decoding process begins with a CNN layer to transform $\hat{Y}$, followed by another CNN layer that increases the feature dimension to 192. Two ResNet blocks and an Attention Block are then used to recover feature dimension. Next, the decoder utilizes four ResNet blocks and UpSample layers to progressively double the image's height $H$ and width $W$ while halving the feature dimension $C$ in each cycle, thereby restoring the image size using features from various dimensions. The UpSample layers are implemented using CNNs. Finally, the decoder employs two ResNet blocks to refine the image based on the 192-dimensional feature information. A CNN layer reconstructs the normalized image with three feature dimensions, and a Tanh activation function $\text{Tanh}(t) = \frac{e^t - e^{-t}}{e^t + e^{-t}}$ generates the final output. Since the output consists of normalized pixel values, a de-normalization layer converts these values back to the [0, 255] range, resulting in the received image $\hat{X}$.

The semantic communication model primarily consists of ResNet Blocks and Attention Blocks. A ResNet Block is composed of two identical structures, each including a Group Normalization (GN) layer, a Swish activation function, and a CNN layer. The GN layer normalizes features within each group to enhance training stability. In our experiments, the number of images per group is set to 32. The Swish activation function, defined as $\text{Swish}(t) = \frac{t}{1 + e^{-t}}$, provides an adaptive non-linear transformation that improves performance and training stability of ANNs. The CNN layer adjusts the feature dimension $C$. In the Attention Block, the input is first normalized by the GN layer and then passed through three CNN layers to produce the outputs \( Q \), \( K \), and \( V \). These outputs are used in the self-attention function \(\text{Attention}(Q, K, V) = \text{softmax}\left(\frac{Q K^\top}{\sqrt{d_k}}\right) V \) to generate new feature vectors, where $d_k$ represents the feature dimension. Finally, a CNN layer transforms these new feature vectors.

\subsection{Training Method}

Using the structure shown in Fig.~\ref{fig:5}, we constructed the semantic communication model. To validate the model, we trained and tested it on the Mini-ImageNet dataset, a subset of ImageNet that consists of 100 classes, each with 600 images, totaling 60,000 images. The training set contains 14,400 images from 64 classes, while the test set includes 12,000 images from 20 classes. The remaining images construct the validation set, which was not used in the experiment. Since the image sizes vary, we cropped all images to [256, 256, 3] for consistency, with a height and width of 256 pixels and 3 feature channels.

It is important to note that there is no overlap between the images or classes in the training and test sets. If the model trained on the training set performs similarly on the test set, it indicates strong generalization capability. Moreover, if the feature distribution parameters obtained from the training set can be applied to the test set for resampling, it would further demonstrate the effectiveness of the VSCC method in extracting semantic features.

We employ three methods for training and testing the model. The first method uses the VSCC loss function Eq.~\ref{eq:10}. The second method utilizes the VAE loss function, which does not account for channel noise. The loss function for the VAE-based semantic communication model is:
\begin{equation}
\begin{aligned}
& \mathcal{L}_{\text{VAE}} = \min \mathbb{E}_{x \sim p(x)} \left[ -\log q(x \mid z) \right. \\
& \quad \left. + \frac{1}{2} \left( -\log \sigma_z^2 + \sigma_z^2 + \mu_z^2 - 1 \right) \right],
\end{aligned}
\label{eq:11}
\end{equation}
where \(\sigma_z\) and \(\mu_z\) represent the variance and mean of the latent variable \(z\), respectively. These parameters are the learned source features obtained from the joint encoder. In the third method, a semantic communication system is modeled as an AE and uses the Mean Squared Error (MSE) as its loss function, as shown below:
\begin{equation}
\mathcal{L}_{\text{AE}} = \frac{1}{H \times W \times C} \sum_{i=1}^{H \times W \times C} (x_i - \hat{x}_i)^2
\label{eq:12}
\end{equation}

In the third method, the end-to-end communication model is constructed without the reparameterization module, enabling direct decoding. Since the reparameterization module does not involve any trainable parameters, the model's complexity remains comparable to that of the VSCC and VAE models.

\begin{algorithm}
\caption{Training Algorithm}
\begin{algorithmic}[1]
\State \textbf{Input:}
\Statex \quad Training set: \( K \)
\Statex \quad Training epoch: \( m \)
\Statex \quad Learning rate: \( \gamma \)
\State \textbf{Initialization:}
\Statex \quad Set of trainable parameters:
\Statex \quad\quad Joint Encoder parameters: \( \theta^{(0)} \)
\Statex \quad\quad Joint Decoder parameters: \( \phi^{(0)} \)
\Statex \quad\quad Encapsulation: \( W^{(0)} = \{ \theta^{(0)}, \phi^{(0)} \} \)
\For{\( t = 0 \) to \( m \)}
    \For{each batch \( \hat{K} \) in \( K \)}
        \State Normalize the input: \( \tilde{x} \leftarrow \text{Normalization}(x) \)
        \State Encode: \( y \leftarrow f_{\text{enc}}(\tilde{x}; \theta^{(0)}) \)
        \State Pass through channel: \( z \leftarrow W(y; \varphi) \)
        \State Decode: \( \hat{x} \leftarrow f_{\text{dec}}(z; \phi^{(0)} ) \)
        \State Denormalize: \( \tilde{\hat{x}} \leftarrow \text{Denormalization}(\hat{x}) \)
    \EndFor
    \If{using the VSCC method}
        \State Compute loss function \( L_{\text{VSCC}} \) using Eq. \ref{eq:10}
    \ElsIf{using the VAE method}
        \State Compute loss function \( L_{\text{VAE}} \) using Eq. \ref{eq:11}
    \ElsIf{using the LAE method}
        \State Compute loss function \( L_{\text{LAE}} \) using Eq. \ref{eq:12}
    \EndIf
    \State Compute gradients: \( \nabla_{W^{(t)}} L \)
    \State Update parameters: \( W^{(t+1)} \leftarrow W^{(t)} - \gamma \nabla_{W^{(t)}} L \)
\EndFor
\State \textbf{Output:} Trained semantic communication model parameters $W$
\end{algorithmic}
\label{al:1}
\end{algorithm}

The models were trained under varying channel SNRs. The selected SNRs were -5, 5, and 15 dB. At very low SNRs, such as -15 dB, the model fails to converge. Conversely, at very high SNRs exceeding 15 dB, the noise variance becomes similar to that of 15 dB, leading to comparable transmission performance, resembling a noise-free channel. To assess the VSCC model's adaptability to different channels, the CMC \(d\) in the Eq.~\ref{eq:10} was adjusted according to training channel SNR. For different training channel SNR, \(d\) values were set to 1, 2, 5, 10 and 15, respectively. The experiment demonstrated that the optimal \( d \) varied across different SNRs, highlighting the VSCC model's ability to adapt to varying channel conditions.

Additionally, to avoid redundant experiments, we adjusted the number of output features \( k \) of the joint encoder as 16. Based on the results from the references~\cite{bourtsoulatze2019deep} and~\cite{kurka2020deepjscc}, the bandwidth compression ratio in our experiment is \(k/n = \frac{16 \times 16 \times 16}{256 \times 256 \times 3} \approx 0.0208\). If the first testing method of VSCC model in section 3.3 is used, the ratio becomes \(k/n = \frac{16 \times 16 \times 32}{256 \times 256 \times 3} \approx 0.0416\), which is significantly lower than the minimum code rate recorded in the references. With traditional separate transmission methods, such as JPEG source coding combined with LDPC channel coding, the image cannot be recovered, as the PSNR falls below 15. Therefore, by selecting 16 as the number of output features, if the image can be successfully transmitted and reconstructed, it indicates that the VSCC method outperforms classical separate transmission methods, making further comparisons with traditional source and channel coding unnecessary. Thus, this study focuses on the similarities and differences of VSCC, VAE and AE methods, and explains the features of VSCC method.

\begin{algorithm}
\caption{Testing Algorithm}
\begin{algorithmic}[1]
\State \textbf{Input:}
\Statex \quad The testing set: \( T \)
\Statex \quad Trained model parameters: \( W = \{ \theta, \varphi \} \)
\Statex \quad Resample number: \( n \)
\Statex \quad Training data variance: \( \sigma_{\text{KB}}^2 \)
\State \textbf{Initialization:}
\Statex \quad Joint Encoder: \( f_\text{en}(\cdot;\theta) \)
\Statex \quad Joint Decoder: \( f_\text{de}(\cdot;\varphi) \)
\For{each batch \( \hat{T} \) in \( T \)}
    \State Normalize the input: \( \tilde{x} \leftarrow \text{Normalization}(x) \)
    \State Encode: \( y \leftarrow f_\text{en}(\tilde{x}; \theta) \)
    \If{AE model}
        \State Pass through channel: \( z \leftarrow W(y; \varphi) \)
        \State Decode: \( \tilde{\hat{x}} \leftarrow f_\text{de}(z; \varphi) \)
        \State Denormalize: \( \hat{x} \leftarrow \text{Denormalization}(\tilde{\hat{x}} \)
        \State Calculate the PSNR using Eq.~\ref{eq:13}
        \State Calculate the SSIM using Eq.~\ref{eq:15}
        \State \textbf{Output:} PSNR, SSIM
    \EndIf
    
    \If{using the VSCC or VAE model}
        \State Get the mean and variance: \( \{\mu_1, \sigma_1^2 \} \leftarrow y \)
        \For{$l=0$ to $n$}
            
            \If{Transmission variance}
                \State Pass through channel: \( z \leftarrow W(y; \varphi) \)
                \State Get: \( \{\mu_1, \sigma_1^2 + \sigma_2^2 \} \leftarrow z \)
                \State Resample: \( \hat{y}^{l} \leftarrow \mathcal{N}(\mu_z, \sigma_1^2 + \sigma_2^2) \)
            \EndIf

            \If{Fixed variance}
                \State Pass through channel: \( z \leftarrow W(\mu_1; \varphi) \)
                \State Resample: \( \hat{y}^{l} \leftarrow \mathcal{N}(z, \sigma_{\text{KB}}^{2}) \)
            \EndIf
            
            \State Decode: \( \tilde{\hat{x}}^{(l)} \leftarrow f_\text{de}(\hat{y}^{(l)}; \varphi) \)
            \State Denormalize:  \( \hat{x}^{(l)} \leftarrow \text{Denormalization}(\tilde{\hat{x}}^{(l)}) \)
            \State Calculate the $\text{PSNR}^{(l)}$ using Eq.~\ref{eq:13}
            \State Calculate the $\text{SSIM}^{(l)}$ using Eq.~\ref{eq:15}
        \EndFor
        \State \textbf{Output:}:
            \State \quad $\text{PSNR}, \text{SSIM}=\mathbb{E} [ \text{PSNR}^{(l)}, \text{SSIM}^{(l)} ]$
            \State \quad $\text{PSNR}_\text{max},\text{SSIM}_\text{max}=\max [ \text{PSNR}^{(l)}, \text{SSIM}^{(l)} ]$
            \State \quad $\text{PSNR}_\text{min},\text{SSIM}_\text{min}=\min [ \text{PSNR}^{(l)}, \text{SSIM}^{(l)} ]$
    \EndIf
\EndFor
\end{algorithmic}
\label{al:2}
\end{algorithm}

It is important to note that, regardless of the modeling approach, the training algorithms remained consistent, as detailed in Algorithm~\ref{al:1}. To maximize the model's feature extraction capability, we experimented with various parameter combinations, balancing transmission performance and computational resource usage. The models were optimized using the Adam optimizer with a learning rate of \(10^{-4}\). Training spanned 200 epochs, with a batch size of 64 due to GPU memory constraints. This is sufficient for the model to converge. All experiments were conducted on a Tesla A100 GPU with about 40 GB of memory.

\begin{figure*}[ht]
    \begin{minipage}{0.9\textwidth}
        \makebox[0pt][l]{\hspace{-0.5cm}\raisebox{0.5cm}{(a)}} 
        \centering
        \includegraphics[width=0.9\textwidth]{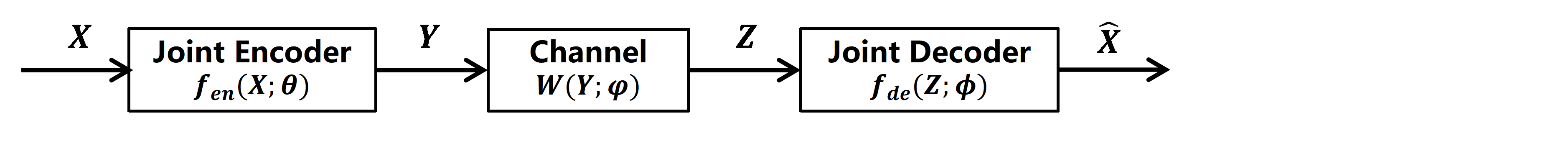}
        \label{fig:6a}
    \end{minipage}

    \vspace{0.5cm} 
    \begin{minipage}{0.9\textwidth}
        \makebox[0pt][l]{\hspace{-0.5cm}\raisebox{1.4cm}{(b)}}
        \centering
        \includegraphics[width=0.9\textwidth]{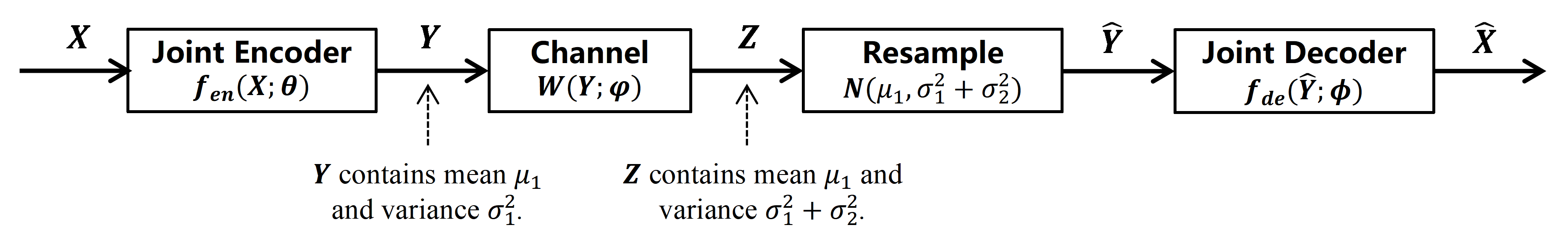}
        \label{fig:6b}
    \end{minipage}

    \vspace{0.5cm}
    \begin{minipage}{0.9\textwidth}
        \makebox[0pt][l]{\hspace{-0.5cm}\raisebox{2.4cm}{(c)}}
        \centering
        \includegraphics[width=0.9\textwidth]{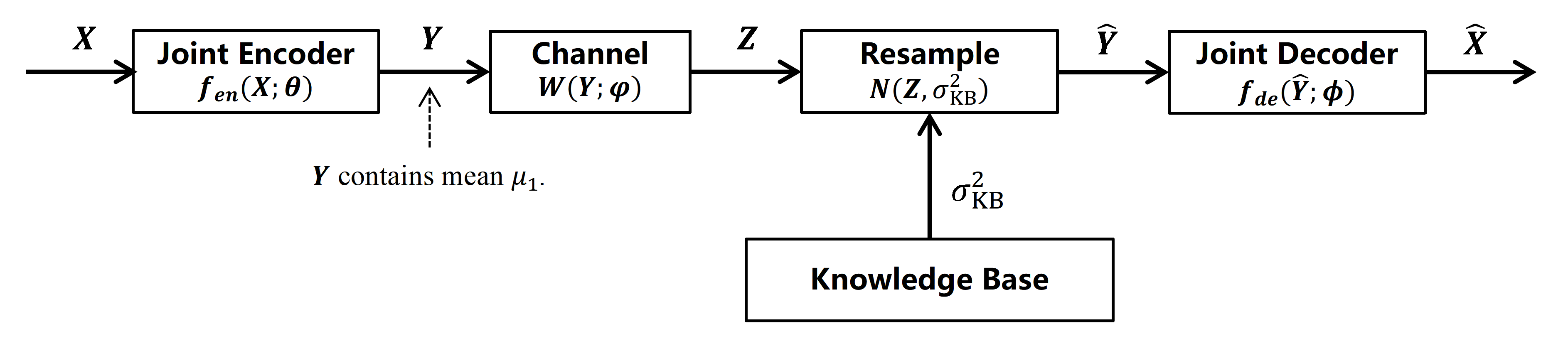}
        \label{fig:6c}
    \end{minipage}

    \caption{Three different testing methods corresponding to Algorithm \ref{al:2}. (a) AE testing method. (b) Transmission variance testing method, where both the mean and variance of the encoded vector \(Y\) are transmitted through the channel. (c) Fixed variance testing method, where only the mean of the encoded vector \(Y\) is transmitted, and the receiver uses a fixed variance from the knowledge base for resampling and decoding.}
    \label{fig:6}
\end{figure*}

\subsection{Testing Method}

Using the DNNs from Section 3.1, three models were trained separately based on Eq.~\ref{eq:10}, Eq.~\ref{eq:11}, and Eq.~\ref{eq:12}. To compare the semantic communication capabilities of the VSCC method with those of the AE and VAE methods, two distinct testing methods were employed, as outlined in Algorithm~\ref{al:2}.

As shown in Fig.~\ref{fig:6}a, for the model trained using the AE method, the image \(X\) is encoded by the encoder and then directly transmitted through the channel. The received data is the latent variable \(Z\), which is directly decoded by the decoder to obtain the reconstructed image \(\hat{X}\). Since the AE method essentially seeks to maximize the likelihood of the reconstructed data, its primary goal is to achieve lossless communication. The channel is implicitly modeled, making it difficult to explicitly understand the role of the channel in the communication process.

For the model trained using the VSCC method, two testing approaches can be employed. Since the VAE method has a similar modeling approach to VSCC, it can be tested using the same methods as VSCC. The first approach, as shown in Fig.\ref{fig:6}b, involves encoding the image \(X\) to obtain the encoded vector \( Y \), which can be regarded as following the distribution $\mathcal{N}(\mu_1, \sigma_1^2)$. It should be noted that \(\sigma_1\) is also learned by the joint encoder and needs to be transmitted to the receiver through the channel. After \( Y \) is transmitted through the channel, the latent variable \(Z\) is received, which includes the mean \(\mu_1\) but with a variance of \(\sigma_1^2 + \sigma_2^2\). Instead of directly decoding \( Z \), the receiver samples from the distribution \(\mathcal{N}(\mu_1, \sigma_1^2 + \sigma_2^2)\) to obtain \(\hat{Y}\), which is then fed into the decoder to reconstruct \(\hat{X}\). The second testing approach, as shown in Fig.~\ref{fig:6}c, differs in that the channel only transmits the mean \(\mu_1\) of the encoded vector \( Y \). The subsequent \(\hat{Y}\) is obtained by sampling from the distribution \(\mathcal{N}(\mu_1, \sigma^2_{\text{KB}})\), where \( \sigma^2_{\text{KB}}\) represents a fixed variance. This fixed variance \( \sigma^2_{\text{KB}}\) is the mean of the variances of the encoded vectors from all training images.

It is important to note that if the first testing approach is chosen, the VSCC method will require transmitting twice the amount of data. In contrast, with the second testing approach, the variance can be stored locally in a knowledge base, eliminating the need for transmission. Since the training and testing data belong to different categories, the variance, which reflects the dispersion of the data, plays a crucial role in determining the optimal data distortion efficiency for different feature distributions under specific channel SNRs. Thus, the VSCC method not only offers a physical explanation for how channel-induced data distortion occurs, but also highlights that the core aspect of semantic extraction using ANNs lies in the variation of data distribution variance. The mathematical differences between these two testing methods will be discussed in detail in Section V.

Additionally, trained models are assessed under varying channel SNRs, ranging from -10 to 25 dB. The evaluation metrics employed the Peak Signal-to-Noise Ratio (PSNR) and the Structural Similarity Index (SSIM) between the original and received images. PSNR, closely associated with the MSE, is a widely used metric in semantic communication systems. It is computed as follows:
\begin{equation}
\text{PSNR} = 10 \cdot \log_{10} \left(\frac{L^2}{\text{MSE}}\right),
\label{eq:13}
\end{equation}
\begin{equation}
\text{MSE} = \frac{1}{n} \sum_{i=1}^{n} (x_i - \hat{x}_i)^2,
\label{eq:14}
\end{equation}
where $L$ represents the maximum pixel value, which is 255 in this case. And $n$ represents the number of pixels.

Given that the AE model utilizes the MSE loss function, the PSNR for the semantic communication model trained with AE method is theoretically expected to be the highest among the three methods. However, PSNR primarily measures the accuracy of data reconstruction, making it more suitable for evaluating lossless transmission. The metric that emphasizes absolute error may not be suitable for semantic communication systems. To address this limitation, we introduced SSIM as an additional evaluation metric. Unlike PSNR, SSIM is a perception-based metric that better aligns with human visual perception. It assesses image quality by considering structural changes, taking into account factors such as luminance masking, contrast masking, and other perceptual phenomena. SSIM can be expressed as:
\begin{equation}
\text{SSIM}(X, \hat{X}) = [l(X, \hat{X})]^{\alpha} \cdot [c(X, \hat{X})]^{\beta} \cdot [s(X, \hat{X})]^{\gamma},
\label{eq:15}
\end{equation}
where \( l(X, \hat{X}) \) represents the luminance difference between two images. Luminance is defined as the average grayscale value of the pixels. It is determined by the following equation:
\begin{equation}
l(X, \hat{X}) = \frac{2\mu_{X} \mu_{\hat{X}} + c_1}{\mu_{X}^2 + \mu_{\hat{X}}^2 + c_1}.
\label{eq:16}
\end{equation}
\( c(X, \hat{X}) \) represents the contrast difference between the two images, capturing how pixels deviate from the mean in terms of contrast:
\begin{equation}
c(X, \hat{X}) = \frac{2\sigma_X \sigma_{\hat{X}} + c_2}{\sigma_X^2 + \sigma_{\hat{X}}^2 + c_2}.
\end{equation}
\( s(X, \hat{X}) \) represents the structural difference between the two images, defined by the linear correlation of pixels:
\begin{equation}
s(X, \hat{X}) = \frac{\sigma_{X\hat{X}} + c_3}{\sigma_X \sigma_{\hat{X}} + c_3}.
\end{equation}
\( c_1, c_2, c_3 \) are constants introduced to prevent numerical instability caused by very small denominators. It should be noted that although SSIM reflects human perception of images better than PSNR, it is still a metric designed based on data recovery and cannot fully capture the level of semantic information.

\section{Results}

\begin{figure*}[ht]
    \centering
    \begin{subfigure}[b]{0.32\textwidth}
        \centering
        \includegraphics[width=\textwidth]{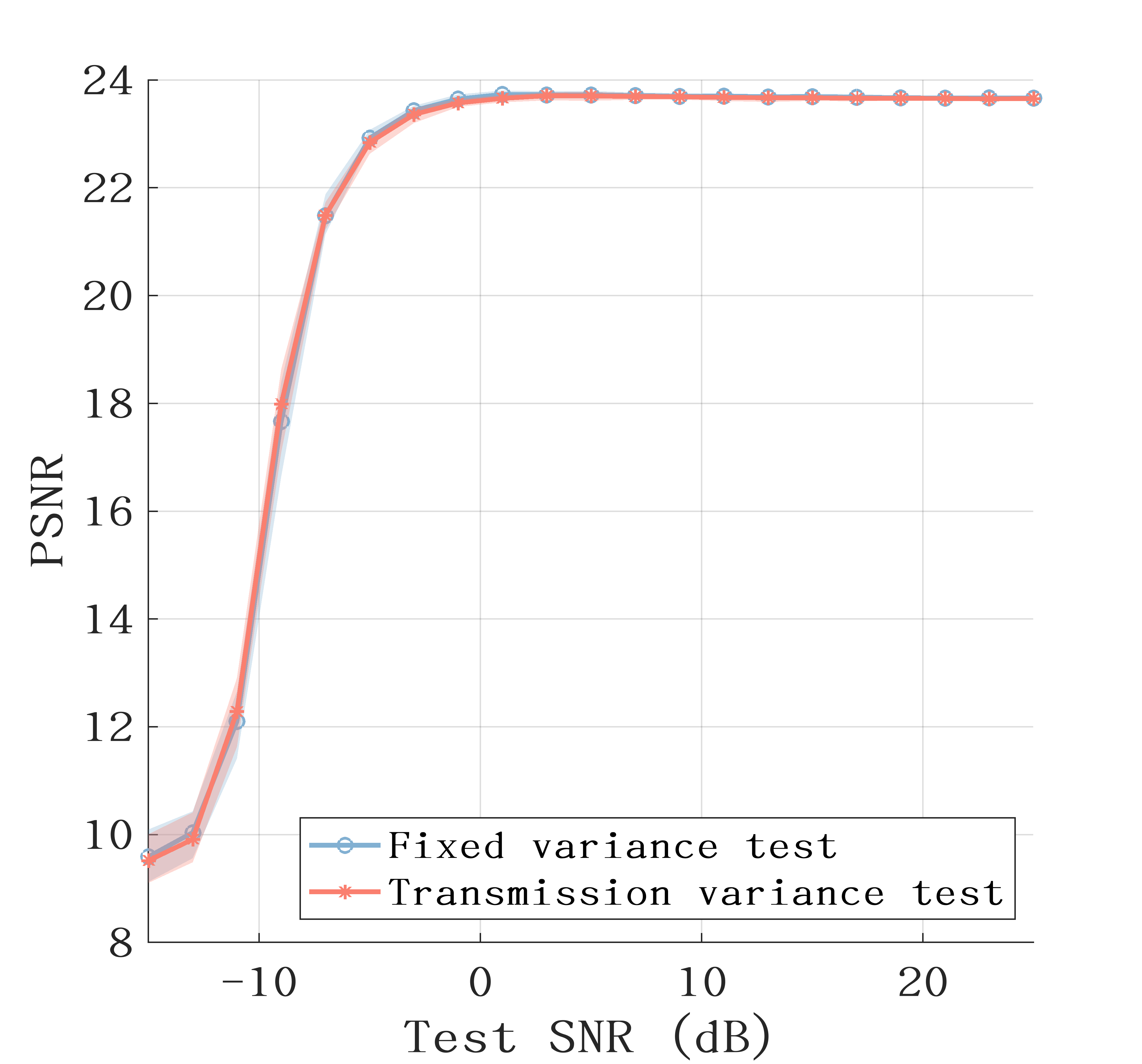}
        \caption{}
        \label{fig:7a}
    \end{subfigure}
    \begin{subfigure}[b]{0.31\textwidth}
        \centering
        \includegraphics[width=\textwidth]{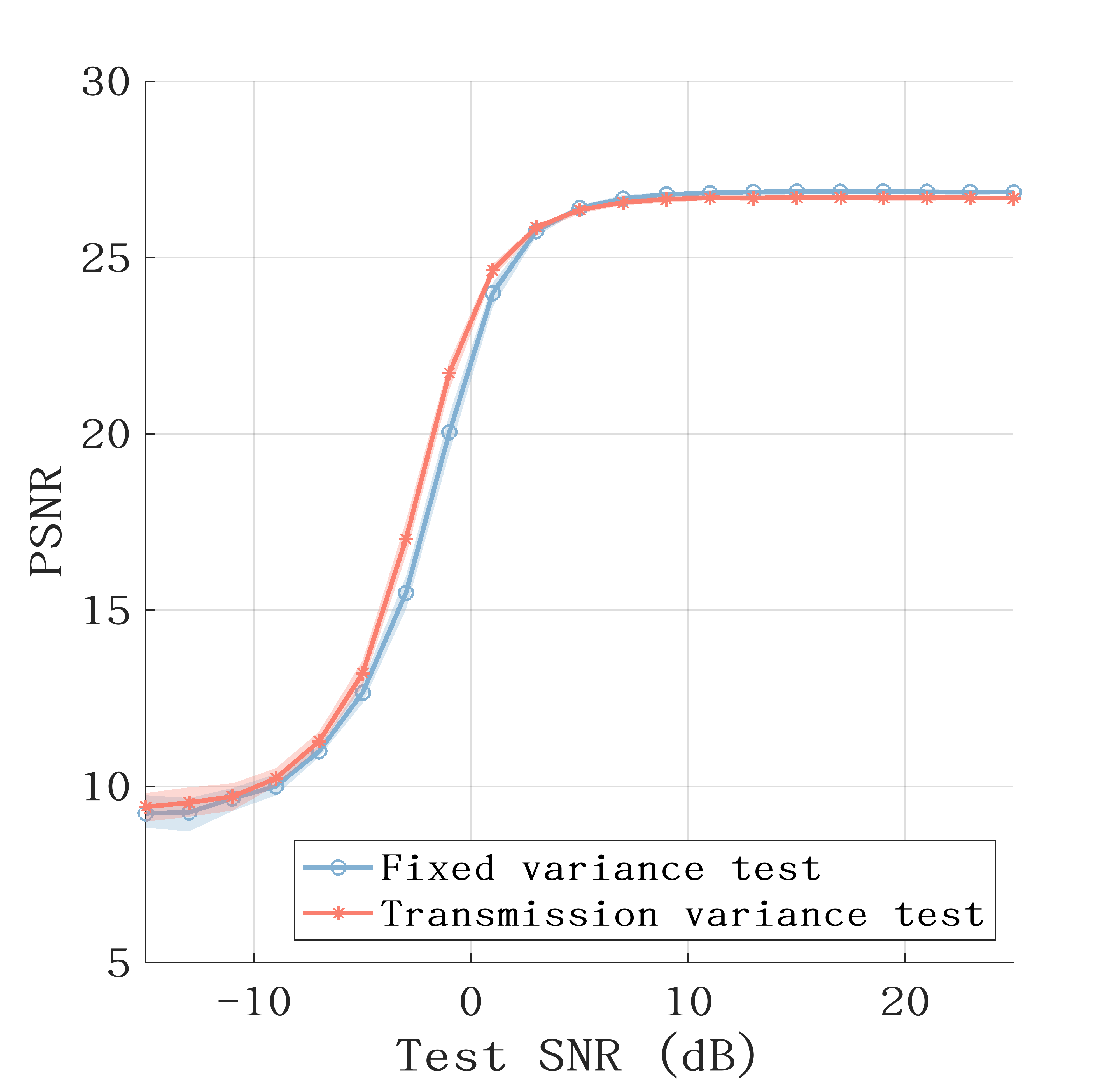}
        \caption{}
        \label{fig:7b}
    \end{subfigure}
    \begin{subfigure}[b]{0.32\textwidth}
        \centering
        \includegraphics[width=\textwidth]{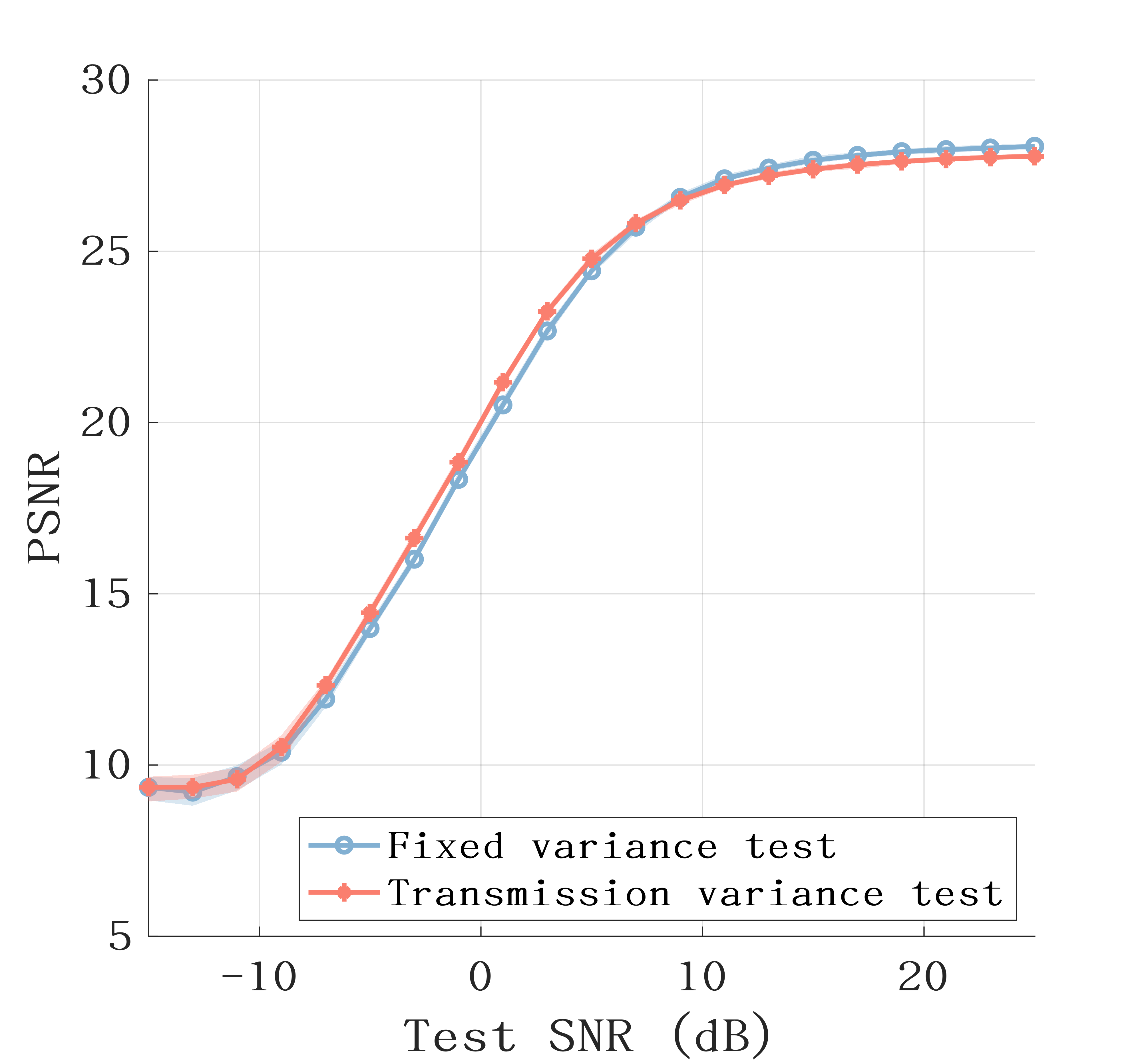}
        \caption{}
        \label{fig:7c}
    \end{subfigure}

    \begin{subfigure}[b]{0.33\textwidth}
        \centering
        \includegraphics[width=\textwidth]{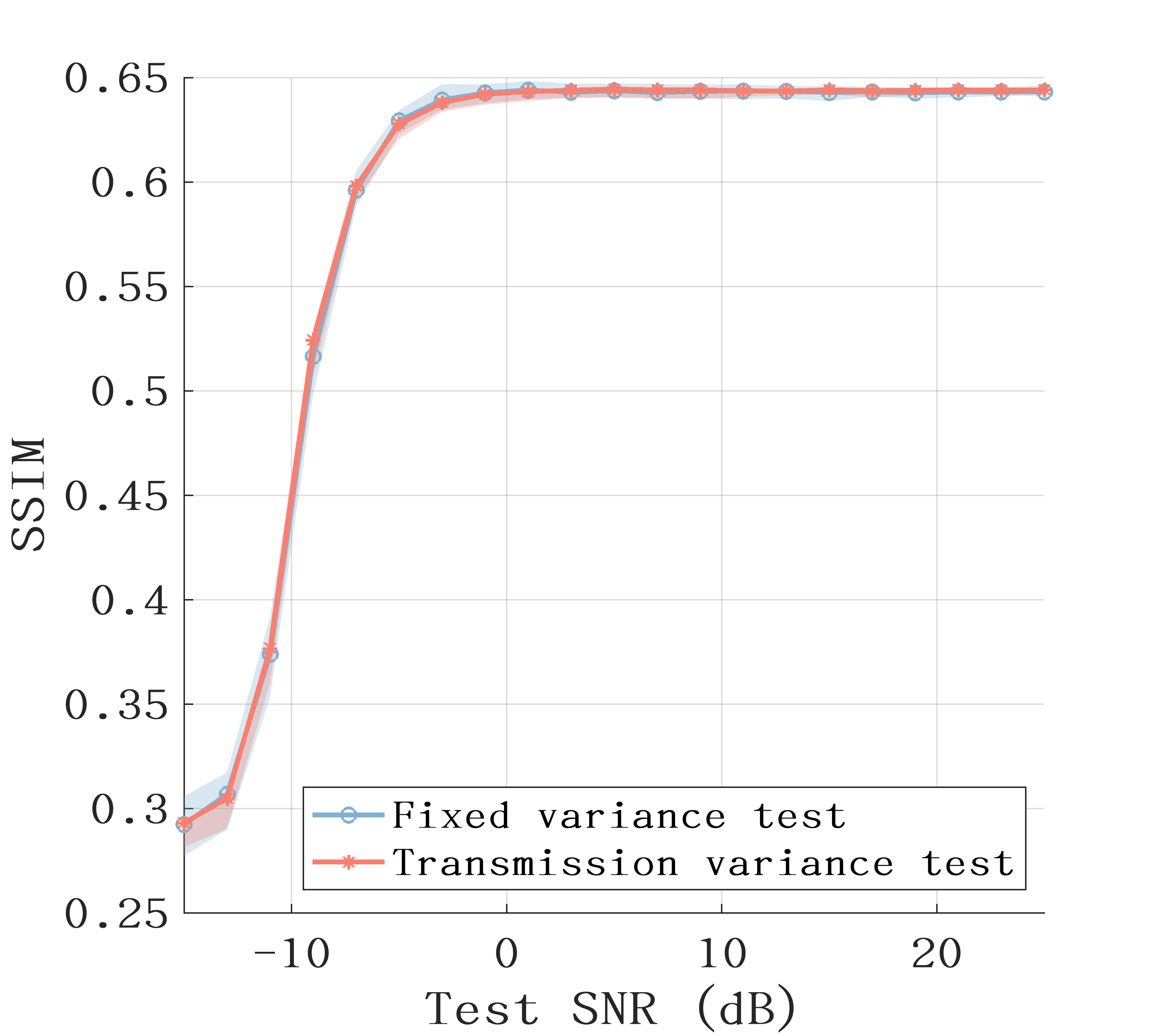}
        \caption{}
        \label{fig:7d}
    \end{subfigure}
    \begin{subfigure}[b]{0.31\textwidth}
        \centering
        \includegraphics[width=\textwidth]{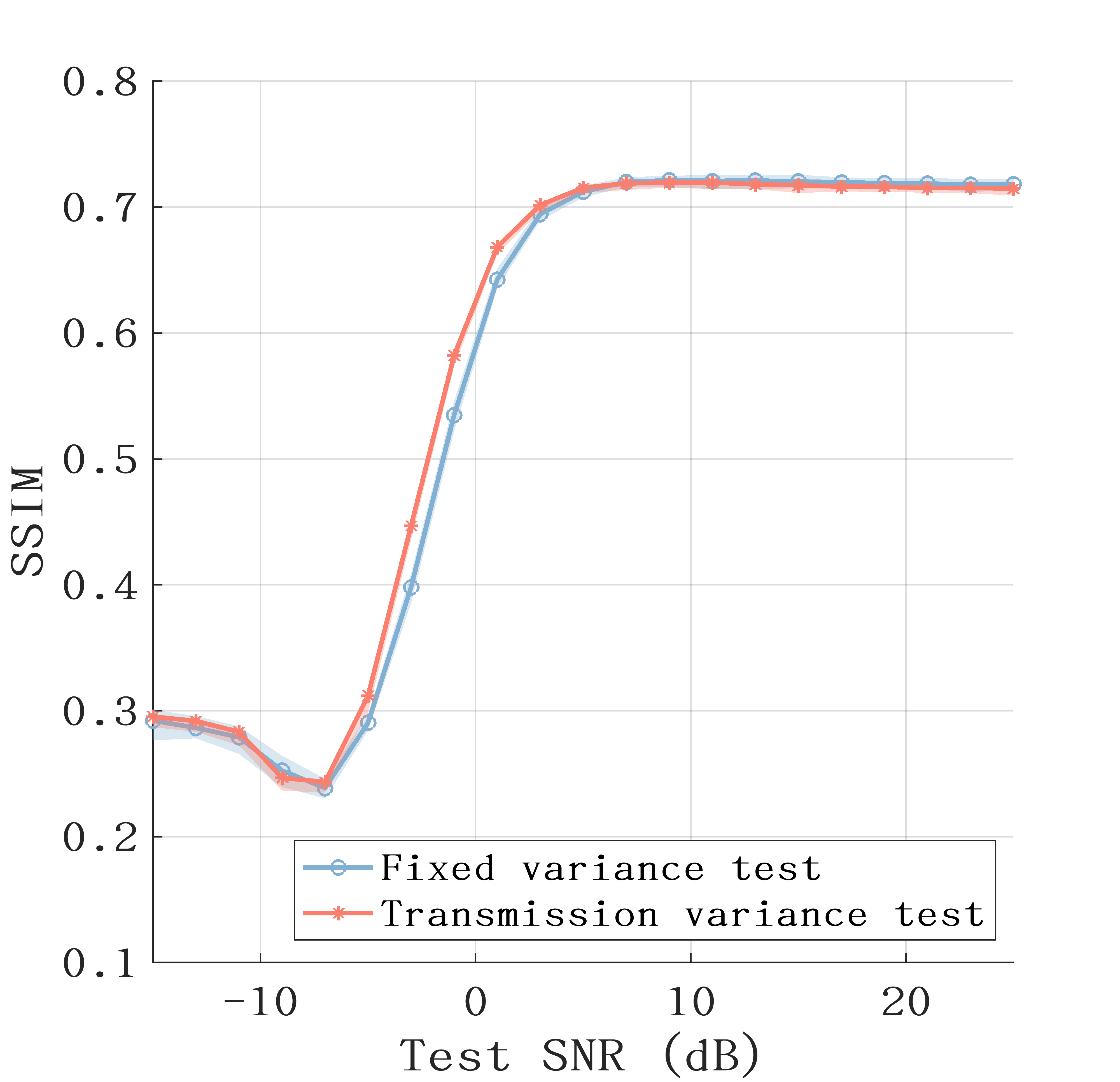}
        \caption{}
        \label{fig:7e}
    \end{subfigure}
    \begin{subfigure}[b]{0.32\textwidth}
        \centering
        \includegraphics[width=\textwidth]{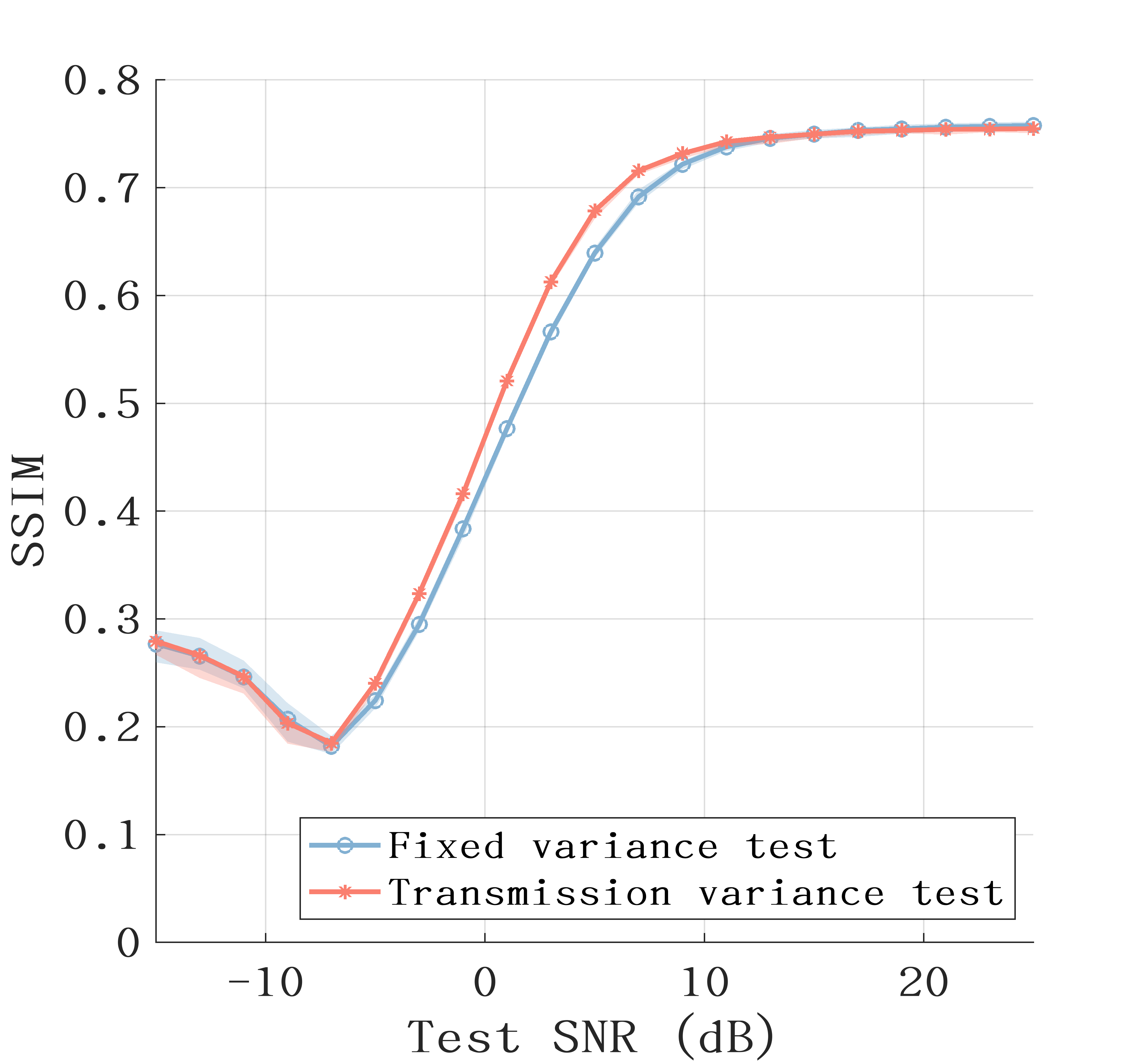}
        \caption{}
        \label{fig:7f}
    \end{subfigure}

    \caption{Experimental results of VSCC models trained under different channel SNRs. These models are tested using two methods. The red line with circle markers represents results obtained using transmission variance resampling, while the blue line with star markers indicates results from fixed variance resampling. Red and blue shaded areas denote the error margins across different samples. (a), (b), and (c) show the transmission performance of VSCC models trained at SNRs of -5 dB, 5 dB, and 15 dB, respectively, with PSNR on the vertical axis indicating data recovery quality. (d), (e), and (f) present the transmission performance under the same SNRs, with SSIM on the vertical axis, which to some extent reflects human perceptual experience.}
    \label{fig:7}
\end{figure*}

In this section, we conducted testing experiments on the VSCC method. The results demonstrate that the VSCC method enables effective semantic communication. The resampling variance during the communication process reflects the distortion effects of the channel. By selecting different CMC $d$ in Eq.~\ref{eq:10} for different training channels, the VSCC model exhibits varying transmission performance. Additionally, we compared the transmission performance of VSCC with that of AE and VAE models. Compared to the AE method, VSCC is a lossy transmission scheme. Its performance may not surpass that of the AE method, but it offers better interpretability, particularly in explaining the role of the channel in JSCC. Compared to VAE, the VSCC method, which incorporates channel parameters directly embedded in the loss function and optimization process, achieves improved transmission performance.

\subsection{Semantic communication capability of VSCC method}

To validate the semantic feature extraction capability of the VSCC method, experiments were conducted using two testing approaches described in Section 3.3. It is important to note that the first testing approach, which involves resampling with transmission variance, requires transmitting twice as much data compared to the second approach, which uses fixed variance. The experimental results indicate that both approaches yield similar transmission performance, suggesting that the variance of the feature distribution, as a measure of data distortion, can be effectively learned by the VSCC model. This also confirms the findings in Section 2.3. The VSCC model learns feature distributions that represent semantics by applying data distortion. It then recovers the original message distribution from these feature distributions, thereby achieving semantic communication.

Since the VSCC method requires resampling for each recovery, 100 resamplings were performed on the same \(z\) for each test, followed by decoding to obtain 100 reconstructed images \(\hat{x}\). The PSNR and SSIM were calculated between each \(\hat{x}\) and the original image \(x\). The maximum, minimum, and average values were recorded. In Fig.~\ref{fig:7}, the red line with circle markers represents the testing results using transmission variance for resampling, while the blue line with star markers indicates the results using fixed variance from the training data. The shaded areas in different colors represent the sampling errors between the extreme values and the average.

\begin{figure*}[ht]
    \centering
    \begin{subfigure}[b]{0.32\textwidth}
        \centering
        \includegraphics[width=\textwidth]{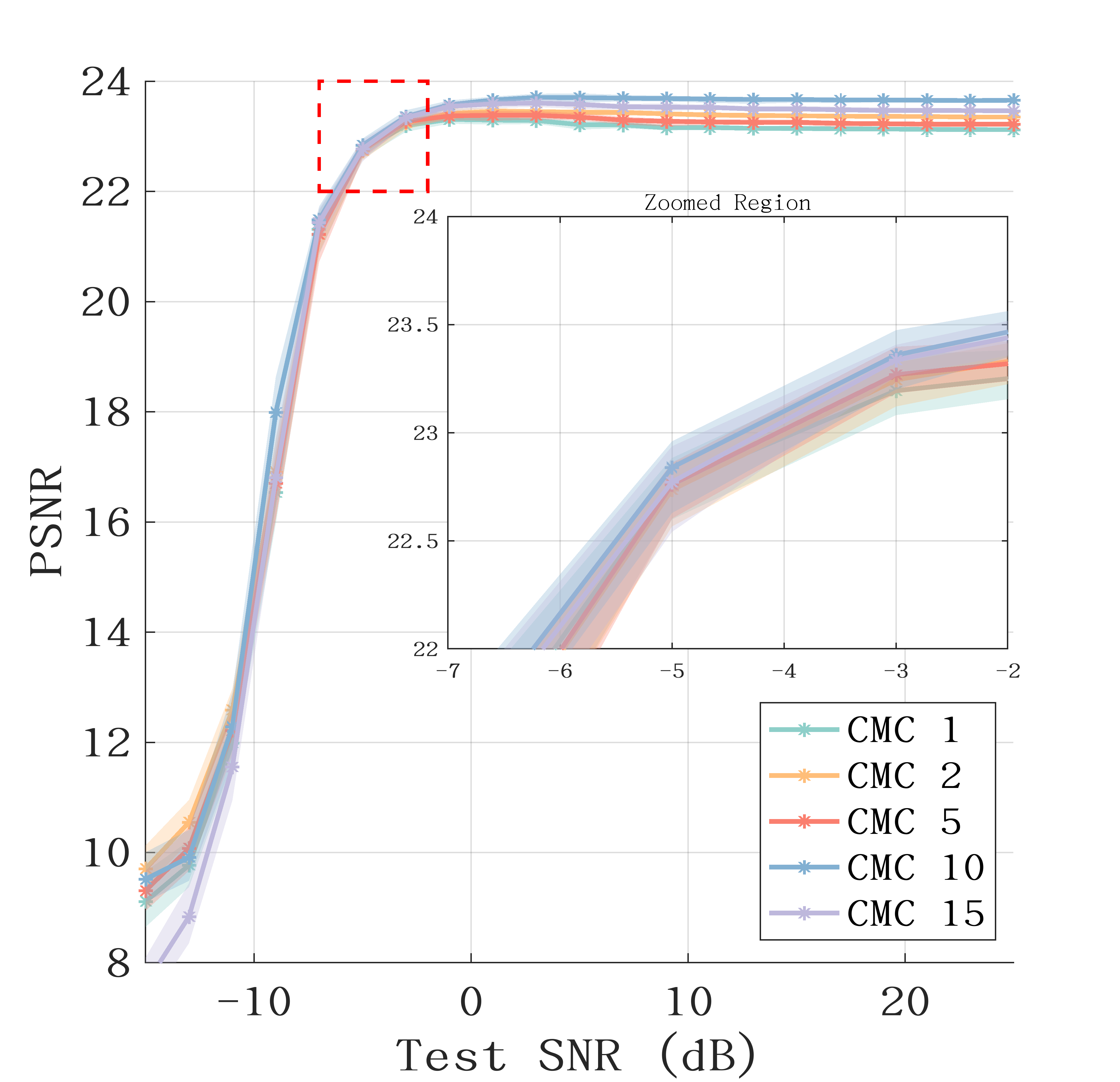}
        \caption{}
        \label{fig:8a}
    \end{subfigure}
    \begin{subfigure}[b]{0.32\textwidth}
        \centering
        \includegraphics[width=\textwidth]{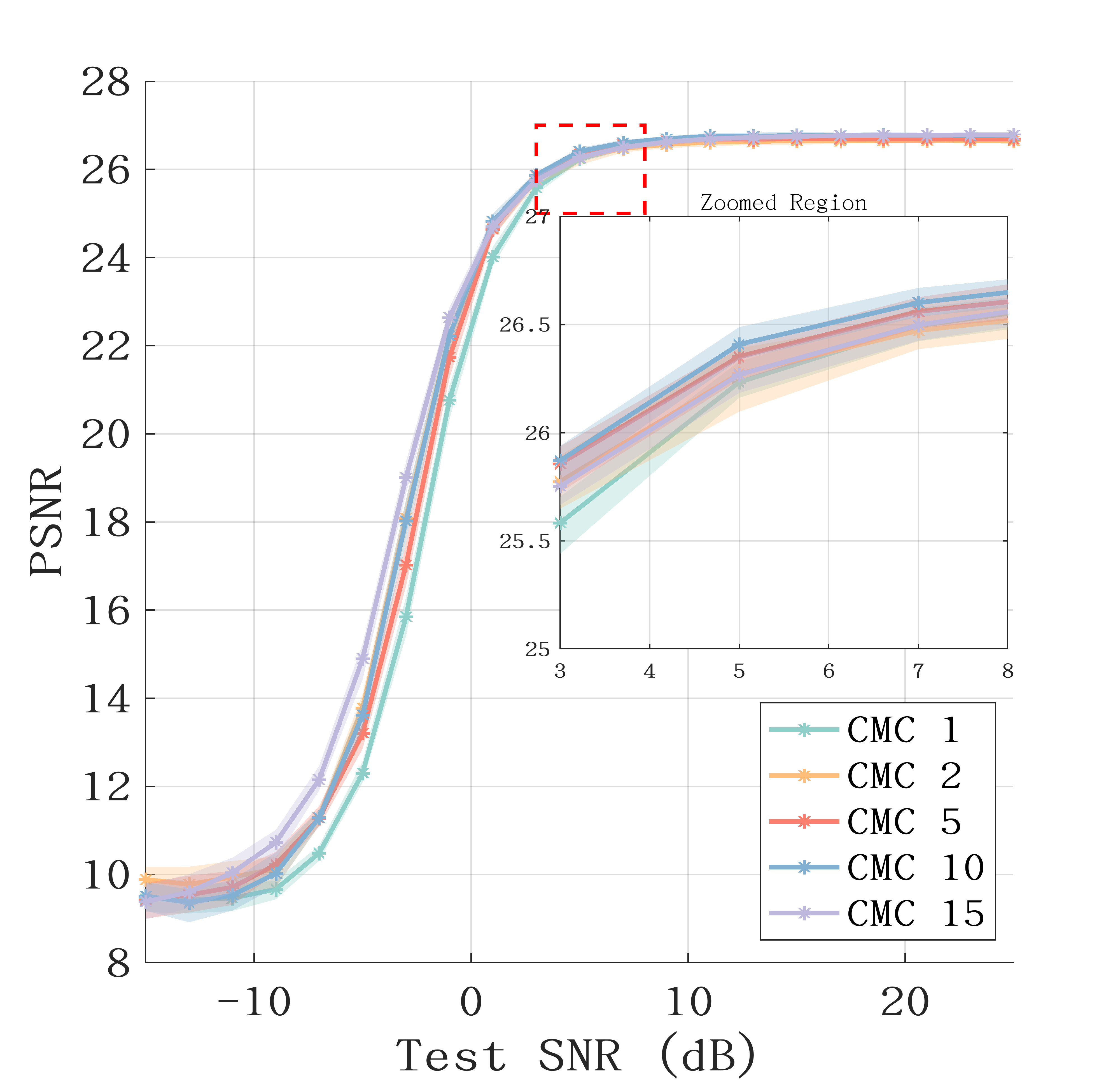}
        \caption{}
        \label{fig:8b}
    \end{subfigure}
    \begin{subfigure}[b]{0.32\textwidth}
        \centering
        \includegraphics[width=\textwidth]{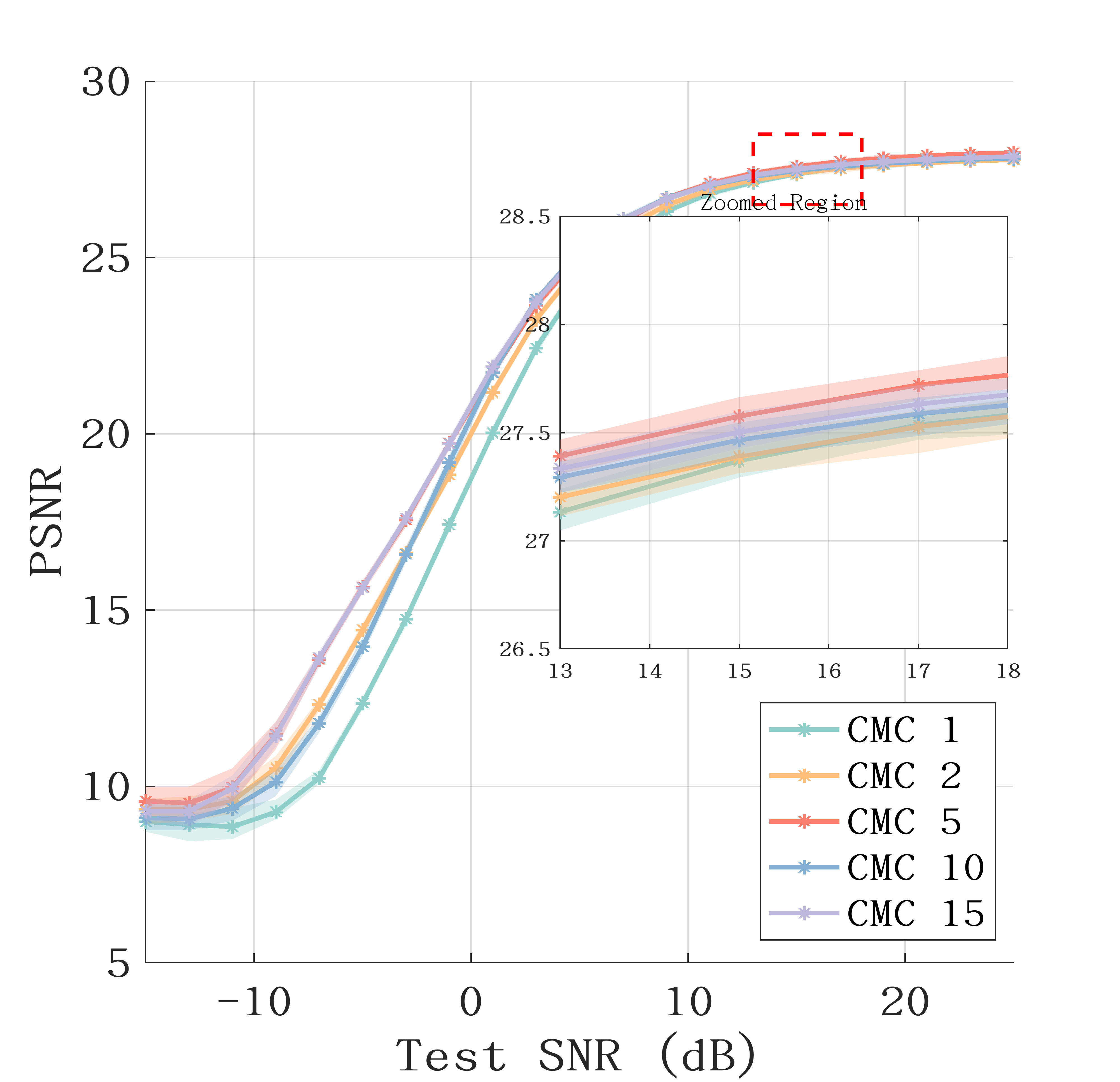}
        \caption{}
        \label{fig:8c}
    \end{subfigure}

    \begin{subfigure}[b]{0.32\textwidth}
        \centering
        \includegraphics[width=\textwidth]{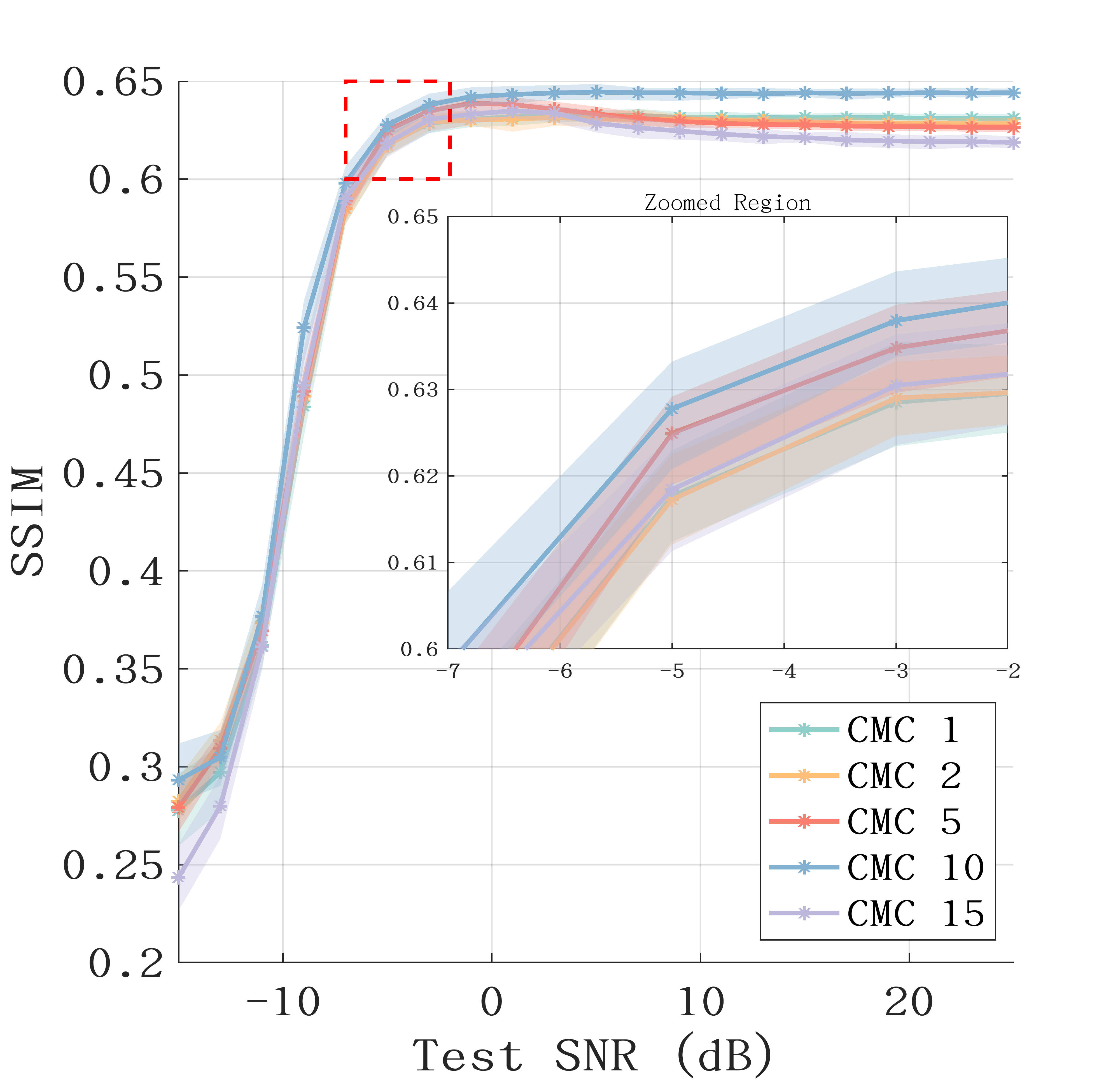}
        \caption{}
        \label{fig:8d}
    \end{subfigure}
    \begin{subfigure}[b]{0.32\textwidth}
        \centering
        \includegraphics[width=\textwidth]{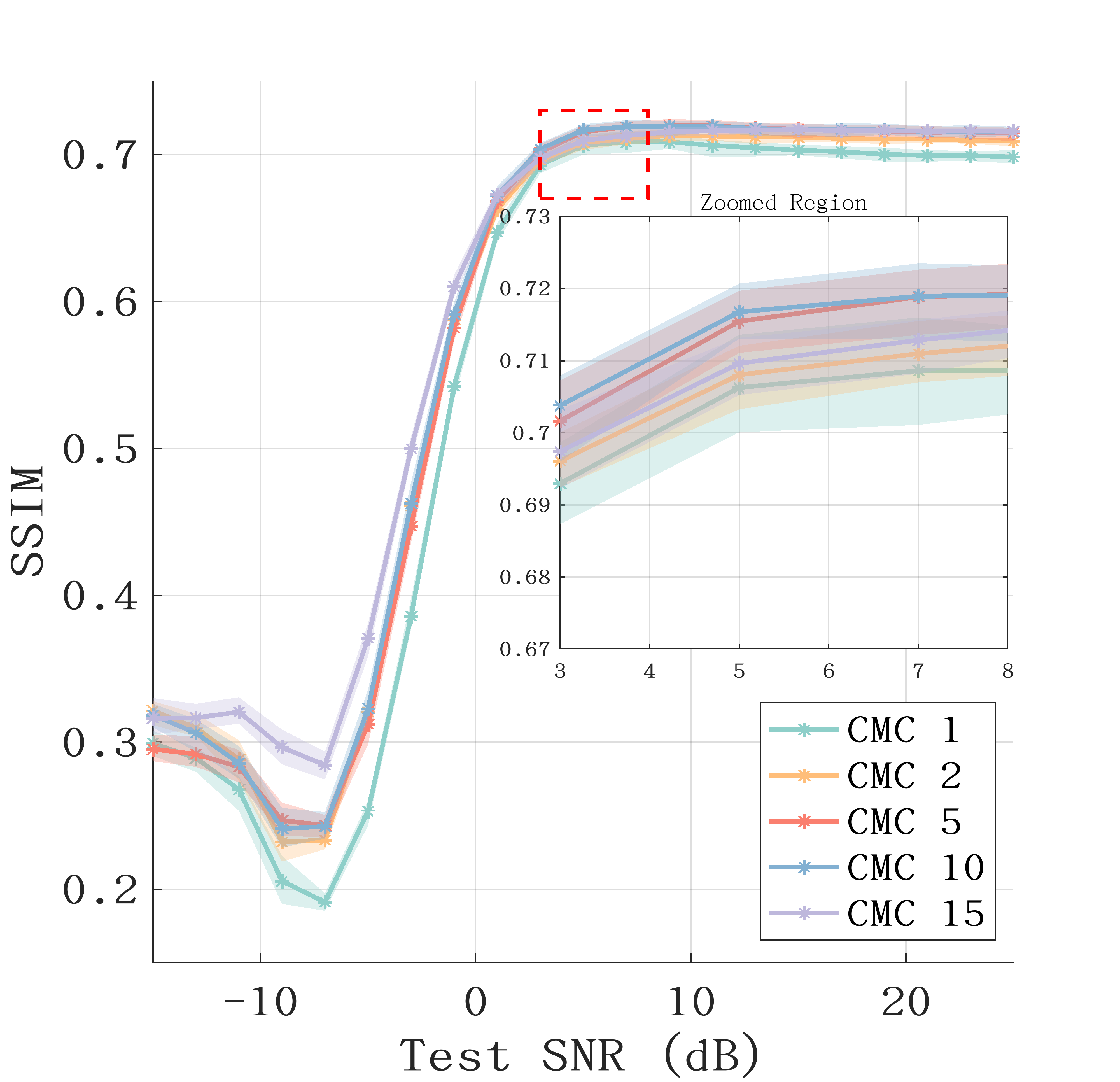}
        \caption{}
        \label{fig:8e}
    \end{subfigure}
    \begin{subfigure}[b]{0.32\textwidth}
        \centering
        \includegraphics[width=\textwidth]{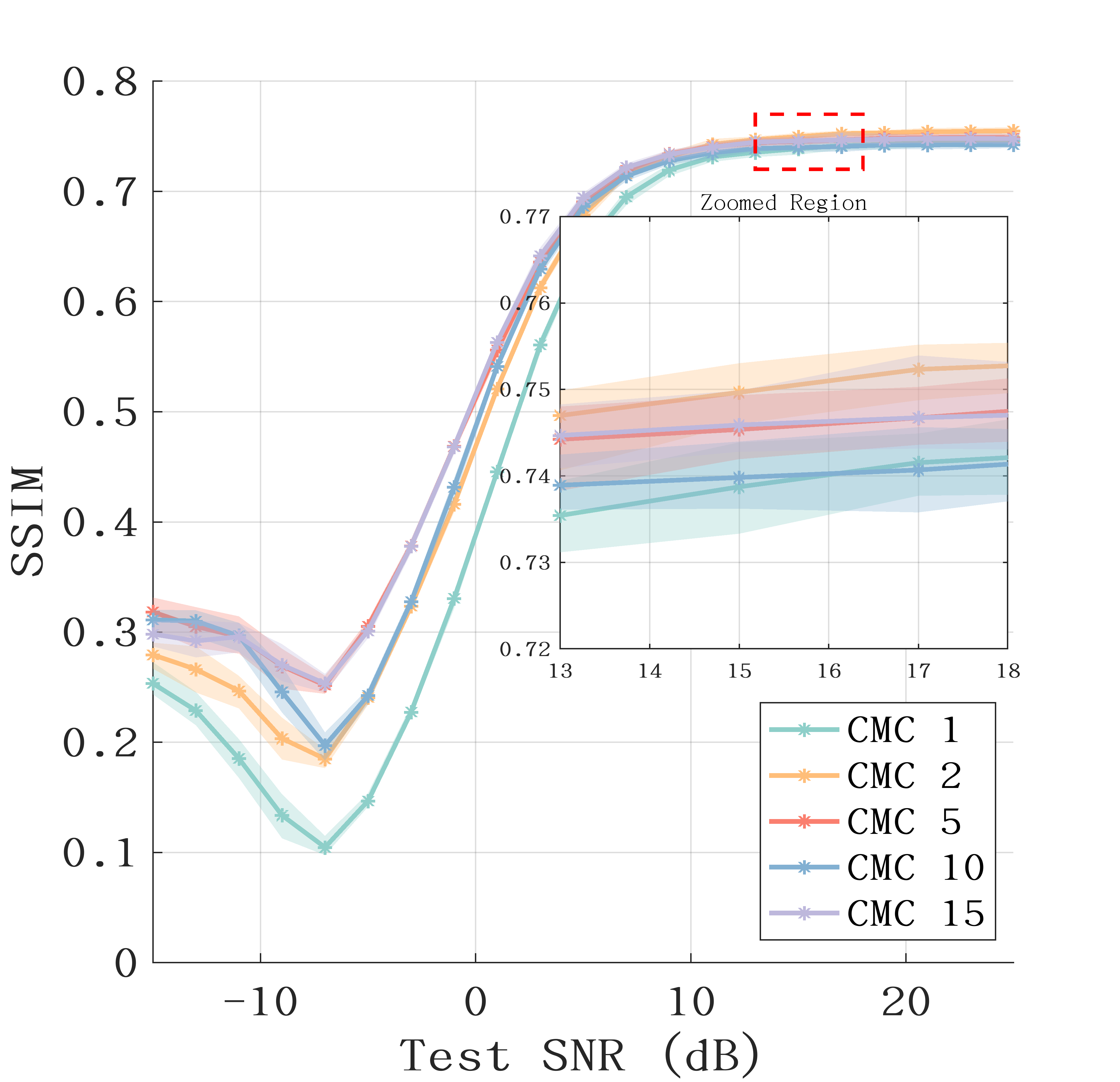}
        \caption{}
        \label{fig:8f}
    \end{subfigure}

    \caption{Experimental results of VSCC models trained with different channel matching constant (CMC) for various channel SNRs, tested using fixed variance resampling. The green, yellow, red, blue, and purple lines represent the testing results of models trained with different CMCs. (a), (b), and (c) show the transmission performance of VSCC models trained at SNRs of -5 dB, 5 dB, and 15 dB, respectively, with the horizontal axis representing the test channel SNR and the vertical axis representing PSNR for data recovery quality. (d), (e), and (f) also display the transmission performance of VSCC models trained at SNRs of -5 dB, 5 dB, and 15 dB, respectively, with SSIM on the vertical axis, which better reflects human perceptual experience compared to PSNR.}
    \label{fig:8}
\end{figure*}

Comparing the red and blue lines in Fig.~\ref{fig:7}, it reveals that the data recovery performance is similar regardless of the testing approach used. In fact, for the model trained at a channel SNR of 15 dB, as shown in Fig.~\ref{fig:7c}, the PSNR obtained from transmission variance resampling are even lower than those from fixed variance resampling. Given that the training and testing data belong to different categories, this indicates that the feature distribution learned by the VSCC method has a certain level of generalization. This generalization is precisely what the semantic encoder aims to capture. However, it is also influenced by the channel and varies with the SNR of the training channel. It is important to note that the VSCC models used in Fig.~\ref{fig:7} were all trained with the optimal CMC. In Section 4.2, we will compare the transmission performance of models trained with different CMC and various channel SNRs.

In addition, comparing Fig.~\ref{fig:7a}, Fig.~\ref{fig:7b}, and Fig.~\ref{fig:7c}, or Fig.~\ref{fig:7d}, Fig.~\ref{fig:7e}, and Fig.~\ref{fig:7f}, it can be found that the transmission capabilities of VSCC models trained under different channel SNRs vary significantly. For example, the model trained at -5 dB achieves a PSNR of 23 and an SSIM of 0.63 when tested at -5 dB. In contrast, the model trained at 10 dB achieves only a PSNR of 13 and an SSIM of 0.3 when tested at -5 dB, while the model trained at 15 dB achieves a PSNR of 13 and an SSIM of 0.22 under the same conditions. This indicates that the current semantic communication models exhibit strong specificity to the channel conditions they were trained on.

Finally, comparing Fig.~\ref{fig:7a} with Fig.~\ref{fig:7d}, Fig.~\ref{fig:7b} with Fig.~\ref{fig:7e}, and Fig.~\ref{fig:7c} with Fig.~\ref{fig:7f}, respectively, it can be observed that better data recovery, as indicated by higher PSNR, does not necessarily correspond to better SSIM. This is particularly evident in Fig.~\ref{fig:7c} and Fig.~\ref{fig:7f}, where the models were trained under favorable channel conditions. Above 10 dB, the PSNR from transmission variance testing is lower than that from fixed variance testing, yet their SSIM values are similar. This discrepancy arises from the different focuses of these two evaluation metrics.

\begin{figure*}[ht]
    \centering
    \begin{subfigure}[b]{0.32\textwidth}
        \centering
        \includegraphics[width=\textwidth]{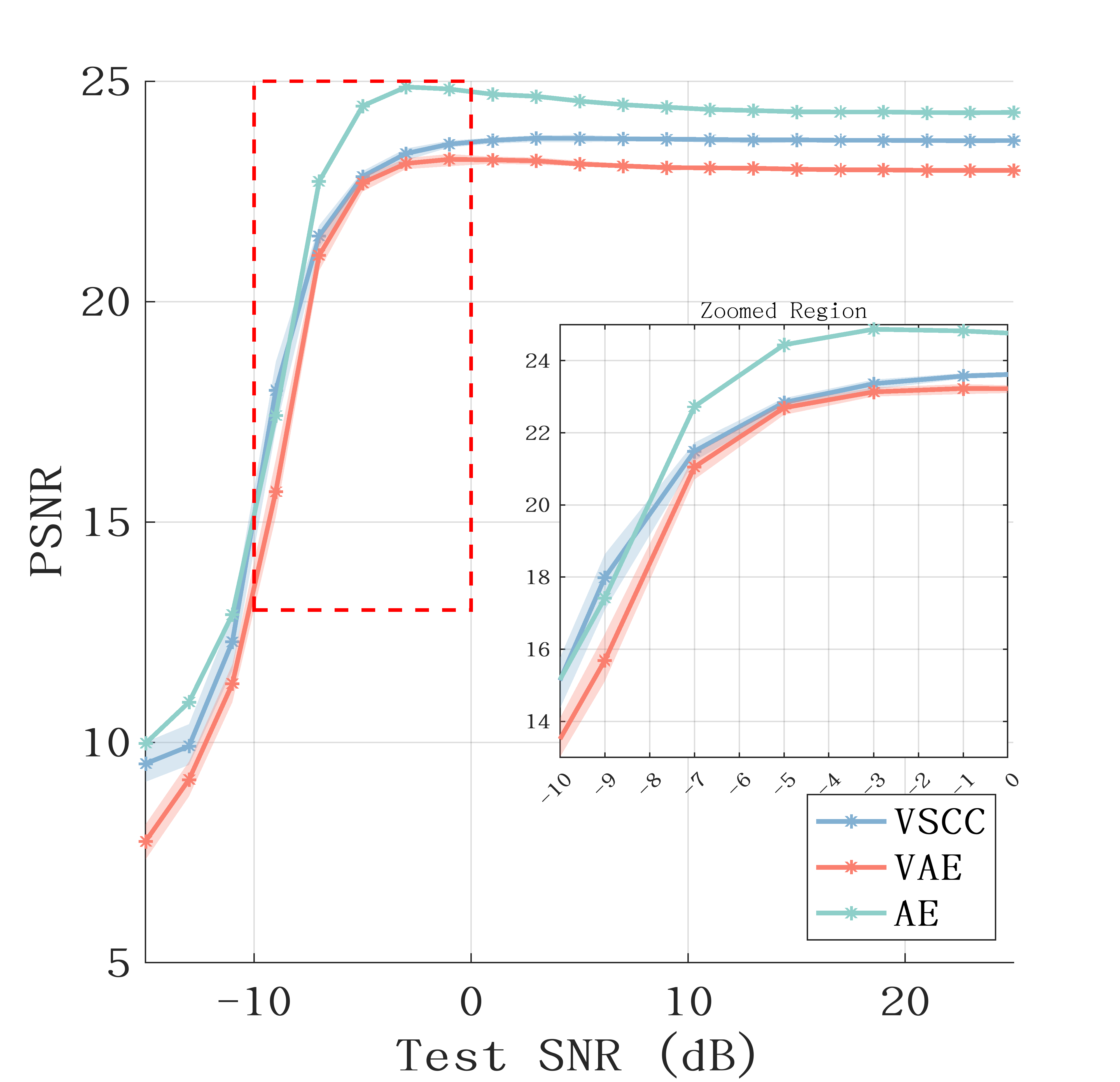}
        \caption{}
        \label{fig:9a}
    \end{subfigure}
    \begin{subfigure}[b]{0.32\textwidth}
        \centering
        \includegraphics[width=\textwidth]{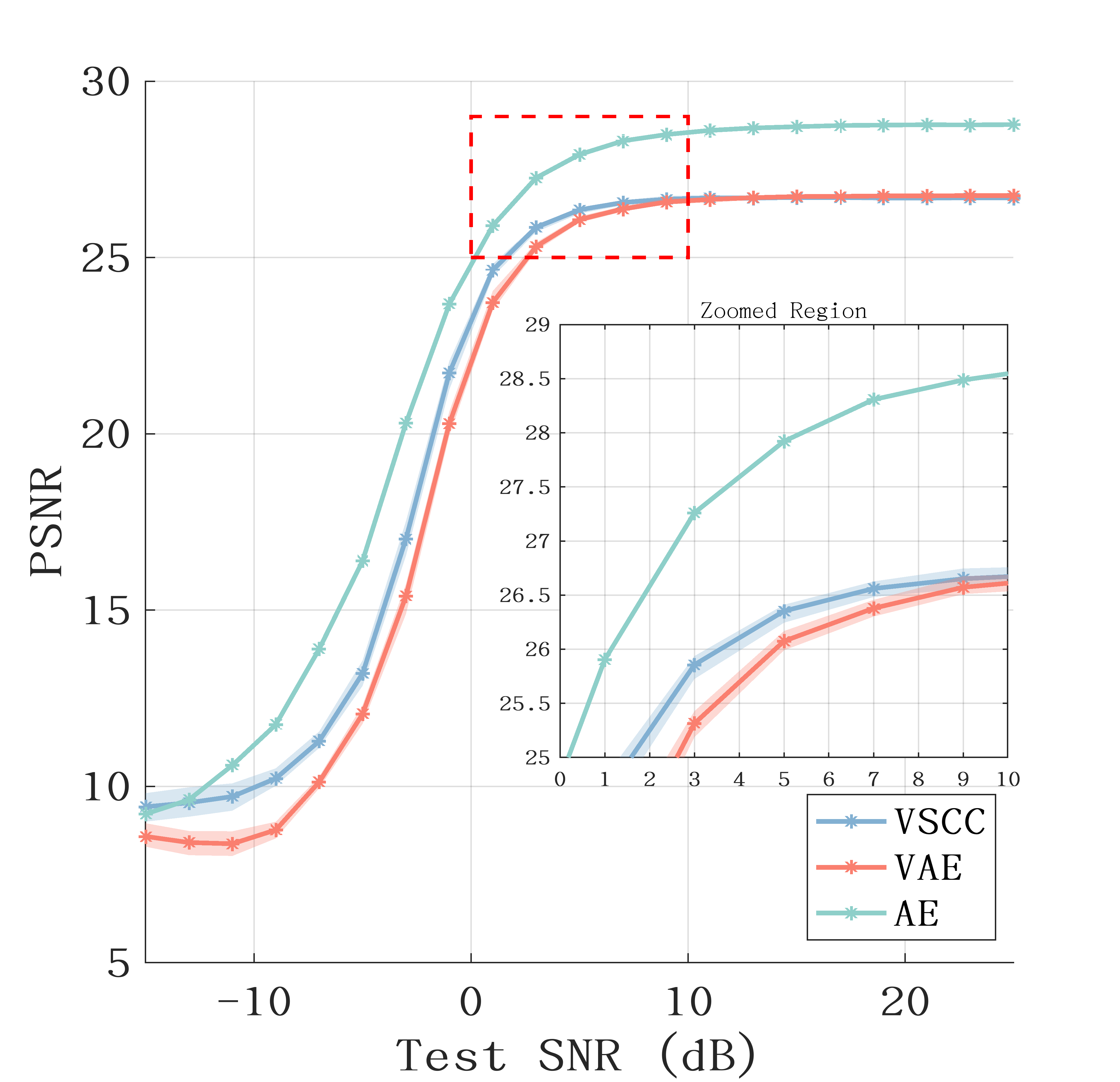}
        \caption{}
        \label{fig:9b}
    \end{subfigure}
    \begin{subfigure}[b]{0.32\textwidth}
        \centering
        \includegraphics[width=\textwidth]{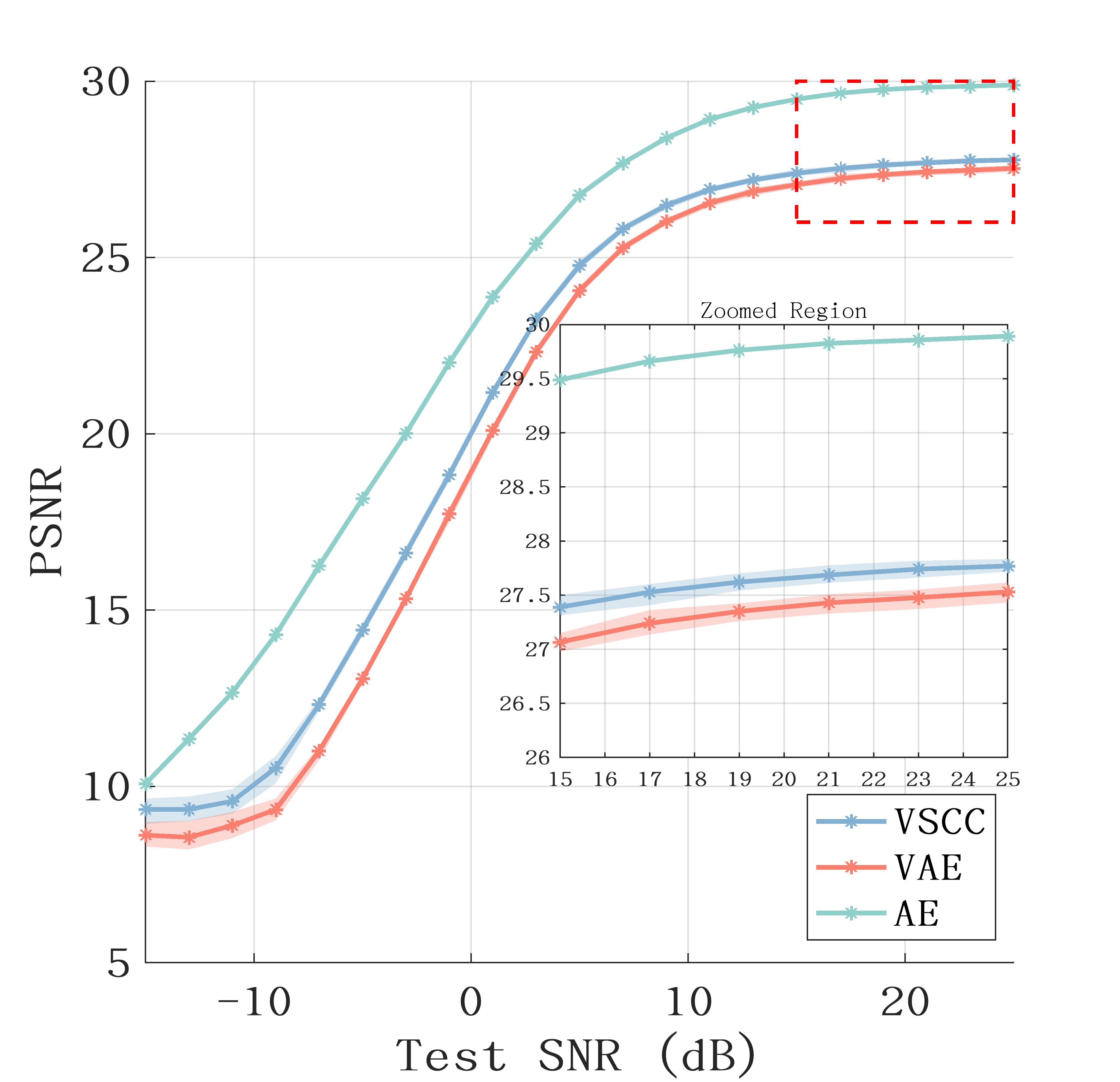}
        \caption{}
        \label{fig:9c}
    \end{subfigure}

    \begin{subfigure}[b]{0.32\textwidth}
        \centering
        \includegraphics[width=\textwidth]{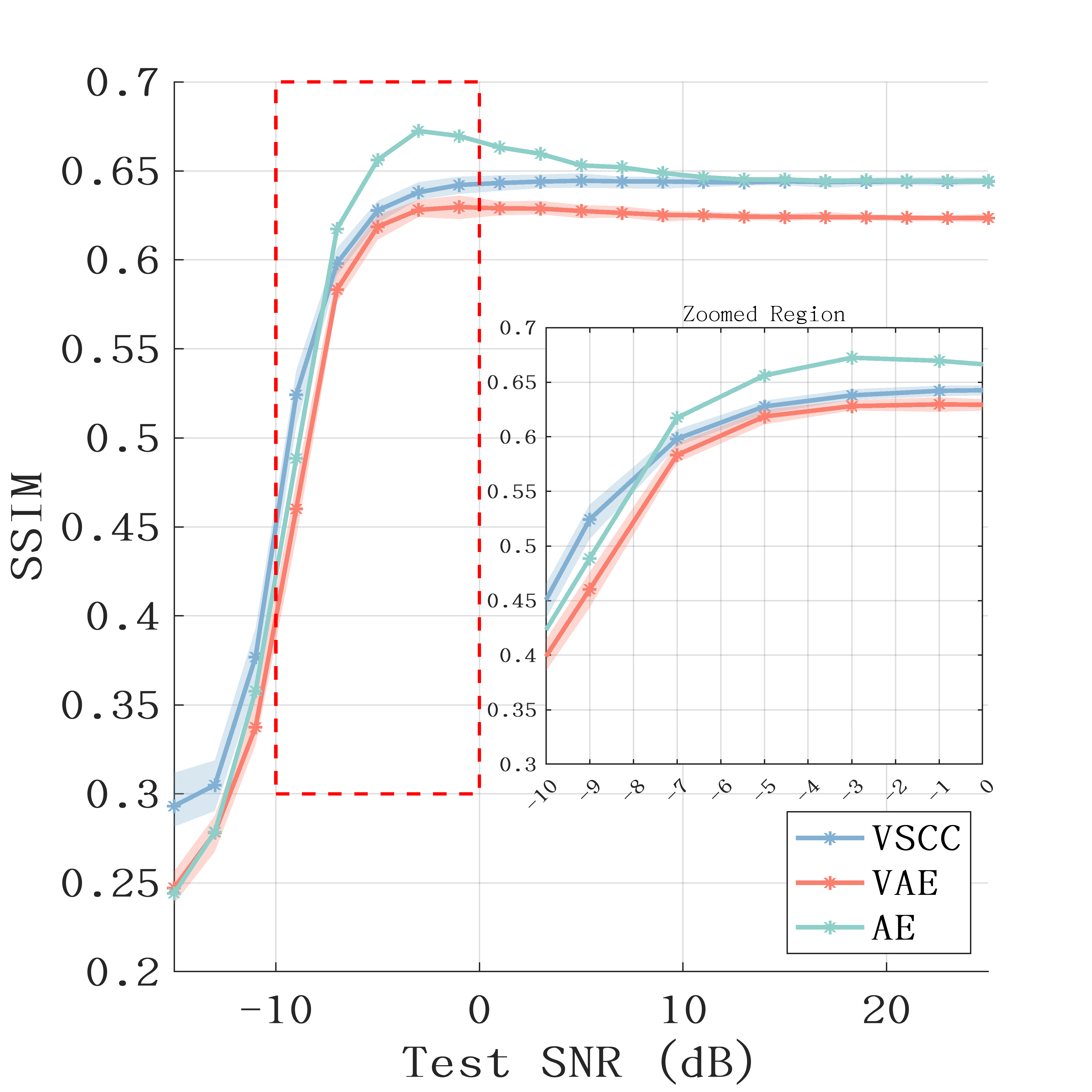}
        \caption{}
        \label{fig:9d}
    \end{subfigure}
    \begin{subfigure}[b]{0.32\textwidth}
        \centering
        \includegraphics[width=\textwidth]{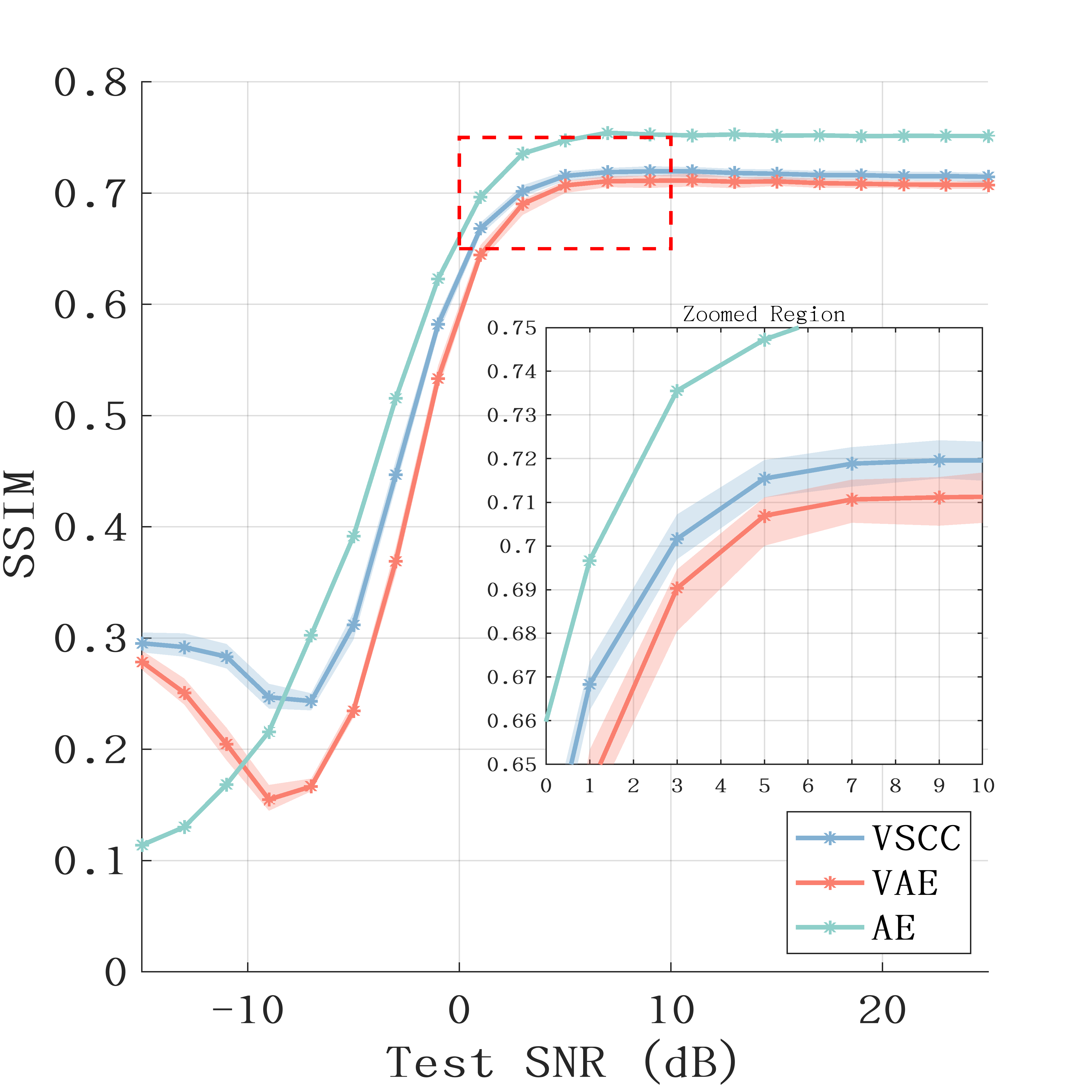}
        \caption{}
        \label{fig:9e}
    \end{subfigure}
    \begin{subfigure}[b]{0.32\textwidth}
        \centering
        \includegraphics[width=\textwidth]{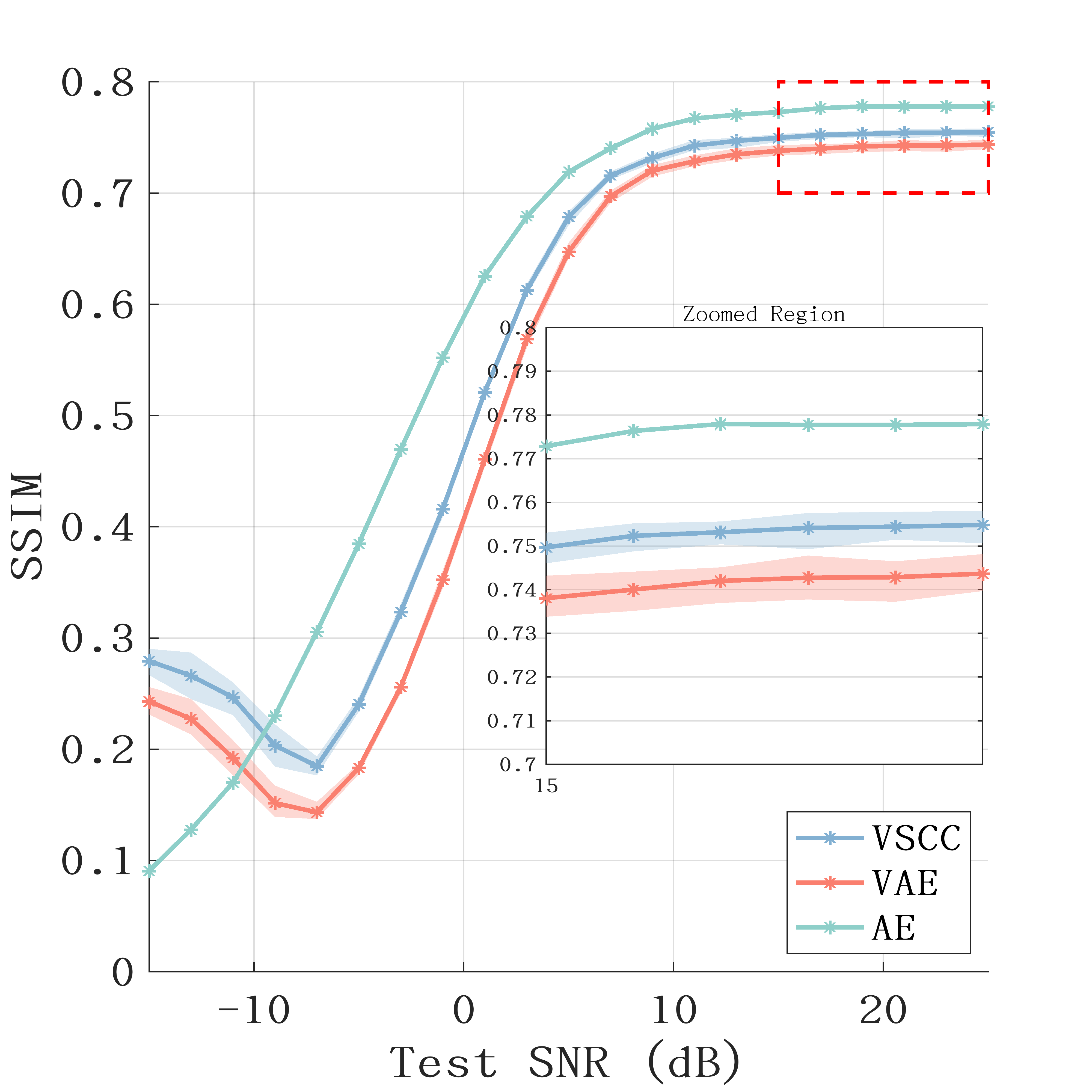}
        \caption{}
        \label{fig:9f}
    \end{subfigure}

    \caption{Experimental results of semantic communication models trained using VSCC, VAE, and AE methods under different channel SNRs, tested with fixed variance resampling. Blue, red, and green lines represent the communication performance of the VSCC, VAE, and AE models, respectively. The shaded areas for the blue and green lines indicate sampling errors for the VSCC and VAE models. (a), (b), and (c) show the transmission performance of models trained at channel SNRs of -5 dB, 5 dB, and 15 dB, respectively, with the horizontal axis representing the test channel SNR and the vertical axis representing PSNR, which indicates data recovery performance. (d), (e), and (f) also show the transmission performance of models trained at SNRs of -5 dB, 5 dB, and 15 dB, respectively, with the vertical axis representing SSIM, a metric that better reflects human perceptual quality compared to PSNR.}
    \label{fig:9}
\end{figure*}

\subsection{Channel matching capability of VSCC method}

In Section 4.1, it can be observed that different transmitted data could be resampled and decoded using the same variance, reflecting a commonality in the feature distributions across different images. The variance represents the degree of dispersion in the data, or in other words, the level of distortion applied to the original data. Such distortion does not impede the semantic communication system's ability to recover the original data distribution. Furthermore, we adjusted the CMC $d$ in Eq.~\ref{eq:10}, which effectively modifies the variance constraint on the encoded vector \(Y\), and training models under different channel SNRs. These models would produce different feature distributions. The analysis in Section 3.1 is further substantiated by the finding that the variance of the feature distribution is correlated with the channel. This correlation arises because the joint encoder in the VSCC model leverages the channel to introduce partial distortion to the data.

As depicted in Fig.~\ref{fig:8}, a comparison of PSNR and SSIM metrics reveals that the semantic communication performance of VSCC models varies depending on the CMC used during training. In Fig.~\ref{fig:8a} and Fig.~\ref{fig:8d}, with a training channel SNR of -5 dB, a CMC of 10 yields the best PSNR and SSIM results across the -5 to 5 dB transmission range. In Fig.~\ref{fig:8b} and Fig.~\ref{fig:8e}, with a training channel SNR of 5 dB, a CMC of 10 provides the best PSNR in the 5 to 15 dB range, while a CMC of 5 achieves the best SSIM. In Fig.~\ref{fig:8c} and Fig.~\ref{fig:8f}, with a training channel SNR of 15 dB, the best PSNR and SSIM results in the 15 to 25 dB range are obtained with CMCs of 5 and 2, respectively. Experimental results for models trained with other CMCs under the same channel conditions are provided in supplementary materials.

The variation of the optimal CMC with different channel SNRs indicates that the channel functions as part of the joint encoder, contributing to data distortion. In Eq.~\ref{eq:9}, the CMC represents the variance constraint on the encoded vector \(Y\) imposed by the joint encoder. Under poor channel conditions, the encoded vector \(Y\) requires greater redundancy to effectively match the channel, which is achieved by increasing the CMC. Conversely, as channel conditions improve, the encoded vector \( Y \) can more closely align with the original data \(X\), resulting in reduced variance. This explains why the optimal CMCs for achieving the best transmission performance, as measured by SSIM, are 10, 5, and 2 for training channel SNRs of -5 dB, 5 dB, and 15 dB, respectively.

Additionally, by comparing Fig.~\ref{fig:8a} with Fig.~\ref{fig:8d}, Fig.~\ref{fig:8b} with Fig.~\ref{fig:8e}, and Fig.~\ref{fig:8c} with Fig.~\ref{fig:8f}, it is evident that the performance of the same model varies when evaluated using PSNR versus SSIM. This further confirms the differing focuses of these two metrics.

It is important to note that under different channel SNRs, the transmission performance of the VSCC model is also influenced by the resampling. In Section 2.3, the loss function Eq.~\ref{eq:10} for the VSCC method includes only a single sampling of the latent variable \(Z\). As a result, when channel conditions are good, using a smaller CMC leads to a lower variance in \(Z\), making it easier to recover the original distribution of \(X\) with just one sampling. This indicates that enhancing the VSCC model with weighted sampling could potentially improve transmission performance. We will explore it further in Section V.

\subsection{Comparison of VSCC, VAE, and AE Methods}

We trained models using the VSCC, VAE and AE methods. The test results of the VSCC, VAE, AE models were compared. The experiment indicates that the AE model delivers the best performance in data recovery. However, the absence of a well-defined semantic evaluation metric makes it difficult to directly assess whether the VSCC model’s semantic recovery capability surpasses that of the AE model. Additionally, the VSCC model outperforms the VAE model.

As shown in Fig.~\ref{fig:9}, the test results of different models are evaluated using PSNR and SSIM. The blue, red, and green lines represent the communication performance of the VSCC, VAE, and AE models, respectively, with the corresponding shaded areas indicating sampling errors. It is important to note that the AE model does not involve a resampling, so there are no resampling errors. The results show that the AE model trained under different SNRs consistently achieves the best performance in both PSNR and SSIM. The VSCC model, due to its channel matching, outperforms the VAE model across all trained SNRs.

On one hand, according to Eq.~\ref{eq:12} in Section 3.2, the AE method is a lossless communication scheme. Although SSIM is generally considered a more reasonable metric than PSNR for assessing image recovery from a human perceptual perspective, it is still fundamentally based on data recovery and cannot fully reflect the effectiveness of semantic-level image restoration. On the other hand, under extreme conditions, as shown in Fig.~\ref{fig:9a} and Fig.~\ref{fig:9d}, the communication performance of the VSCC model in some cases surpasses that of the AE model. Notably, in Fig.~\ref{fig:9d}, the VSCC model achieves the best SSIM performance within the training SNR range, indicating superior recovery. This indirectly demonstrates the semantic communication capability of the VSCC model.

Furthermore, if we consider the recovery performance of the AE model as the upper bound and that of the VAE model as the lower bound, a comparison of Fig.~\ref{fig:9a} with Fig.~\ref{fig:9d}, Fig.~\ref{fig:9b} with Fig.~\ref{fig:9e}, and Fig.~\ref{fig:9c} with Fig.~\ref{fig:9f} reveals that the SSIM recovery performance of the VSCC model is superior to its PSNR recovery performance. This is because the SSIM metric partially evaluates the data distribution rather than solely assessing lossless data recovery. This aligns with the goal of the VSCC model, which is to encode a better feature distribution to achieve the transmission of the original data distribution.

\section{Conclusion and Future Work}

\subsection{Conclusion}

This paper identifies three key challenges in semantic communication. To address the first problem, we highlight that the primary distinction between semantic communication and classical communication lies in whether the data is lossy. Based on research in semantic information, it is clear that while semantics inherently depend on the existence of data, not all data necessarily contribute to semantics. Consequently, the goal of semantic encoding is to eliminate redundant data and extract its semantic features. The semantic communication significantly reduces the amount of data that needs to be transmitted, which is a major source of its gain.

For the second problem, semantic communication extends beyond the boundary conditions of CIT by incorporating a knowledge base and specific communication tasks. This enables JSCC to achieve optimal transmission while further reducing data volume. Building upon the IB principle, we can establish a semantic communication system model within the JSCC framework. Moreover, by implementing VIB, we can integrate deep learning into the practical training process of semantic communication systems. However, the VIB characterizes the process of information compression and task completion from an information-theoretic perspective, yet it does not specify concrete implementation methods for encoding and decoding. Therefore, we introduce variational inference to conduct detailed modeling of the encoder and decoder that incorporate the channel.

At last, we propose an innovative VSCC approach by incorporating channel modeling into the encoder, which simultaneously provides the solution to the third problem. Unlike VAE, which deterministically presumes a predefined distribution for latent variables, VSCC dynamically derives the latent distribution by integrating channel modeling into the encoder. This approach naturally induces channel-adaptive latent distributions, with theoretical analysis revealing their dynamic variation with channel conditions. Furthermore, as a communication-oriented framework, VSCC introduces CMC as a tunable hyperparameter in variational inference to balance channel noise against source distortion. Future work will focus on developing a learnable component within the VIB framework to adaptively control information compression, thereby achieving better channel adaptation. In contrast, VAE remains solely a generative model without such transmission optimization mechanisms.

We constructed the semantic communication system using ResNet Blocks and trained it with three approaches: VSCC, VAE, and AE methods. Experiments show that the VSCC model effectively extracts common features from images, outputting them as the variances of the feature distributions. These variances, representing data distortion, align with lossy transmission. Further, it can be found that the variances of feature distributions are influenced by the training channel SNRs, and varying the CMC during training leads to different transmission performances. It demonstrates that the channel is integrated into the production of feature distributions in the VSCC, becoming part of the joint encoder rather than an obstacle. The VSCC model outperforms the VAE model in data recovery but falls short of the AE model, which targets lossless transmission. While an objective semantic evaluation metric is lacking, the comparison of PSNR and SSIM suggests that the VSCC method achieves a degree of semantic recovery, with transmitted data more closely aligning with human perception.

\subsection{Future Work}

Nevertheless, the question of whether information containing semantic features is necessarily less than the original data information remains open for further investigation. It is generally understood that data correlation does not overlap with the causality of the physical world. Shannon entropy, which is based on the assumption of data independence, compresses data information to its minimum but does not account for causality. However, as discussed in the Introduction, causality can contribute to data correlation in semantics, allowing for further data compression. Furthermore, whether the causality of data is related to the channel is a critical factor in determining whether JSCC is always necessary to achieve optimal performance in semantic communication. Encoding and decoding are fundamentally transformations of data distribution. In particular, when source coding is described by a rate-distortion function, it relaxes the conditions for distribution transformation, allowing the channel to be incorporated. However, the specific conditions under which JSCC achieves optimal performance in various scenarios require further study. For instance, the exact form of Eq.~\ref{eq:2} in Section 2.3 and its relationship with the channel require further investigation. In addition, while data distribution features may capture some semantic characteristics, such as shape and color, they cannot represent all semantic features—such as the difference between a sketch of a mountain and a photograph of a real mountain. The relationship between data distribution and semantic meaning requires deeper exploration to fully understand how these features interact and contribute to the effectiveness of semantic communication.

In our experiments, the VSCC method holds potential for further improvement. In Section 2.3, we assume that the encoded vector \(Y\) follows the distribution \(\mathcal{N}(\mu_1, \sigma_1^2)\), where the mean \(\mu_1\) and variance \(\sigma_1^2\) are learned according to Eq.~\ref{eq:10}. So the variance is constrained by CMC. Nevertheless, in the actual implementation of the VSCC model, after the joint encoder processes the original image \(x\), it outputs the mean \(\mu_x\) and variance \(\sigma_x^2\), which together constitute the encoded vector \(Y\). The mean \(\mu_x\) can be interpreted as a sample drawn from the distribution \(\mathcal{N}(\mu_1, \sigma_1^2)\) and the variance \(\sigma_x^2\) should tend to \(\sigma_1^2\). If the first testing method from Section 3.3 is employed, both the mean \(\mu_x\) and the variance \(\sigma_x^2\) are transmitted. The received latent variable \(Z\) should include the mean \(\mu_z\) and the variance \(\sigma_z^2\) that follow the distributions \(\mathcal{N}(\mu_1, \sigma_1^2 + \sigma_2^2)\) and \(\mathcal{N}(\sigma_1^2, \sigma_2^2)\), respectively. Thus, resampling should be conducted according to the distribution \( \mathcal{N}(\mathcal{N}(\mu_1, \sigma_1^2 + \sigma_2^2), \mathcal{N}(\sigma_1^2, \sigma_2^2)) \), which introduces an unstable variance. There should be a variance estimation process to correct the mean. However, this issue is overlooked in the VSCC loss function Eq.~\ref{eq:10}, leading to a decline in performance. Conversely, if the second testing method is adopted, where only the mean \(\mu_x\) following the distribution \(\mathcal{N}(\mu_1, \sigma_1^2)\) is transmitted, it aligns seamlessly with the derivation of Eq.~\ref{eq:10}. However, because the true variance, following the \( \mathcal{N}(\sigma_1^2, \sigma_2^2) \), is replaced by a fixed variance constrained by CMC, there would also be performance degradation. Consequently, if future work incorporates techniques from Weighted VAE to apply weights for multi-sample latent variables $Z$, where the weights can include the variance estimation process, the loss function of the VSCC method could approximate the maximum a posteriori estimation more closely, thereby enhancing the recovery of both data distribution and the data itself. It could potentially enable the VSCC model to surpass the performance of the AE model in testing. Additionally, the VSCC method illustrates that variance, as an indicator of data dispersion, is crucial for adapting to the channel and extracting shared semantic features from different images in semantic communication. It highlights the importance of estimation for the true variance, which happens to be a key topic in diffusion model research. By incorporating diffusion models to enhance the extraction and recovery of feature distributions represented by latent variables, the performance of the VSCC method could see significant improvement. Furthermore, the VSCC model currently lags behind the AE model when evaluated using existing metrics. This discrepancy does not necessarily indicate the inferiority of the VSCC method compared to the AE method. Instead, it arises from the fact that current evaluation metrics primarily focus on data recovery, which fails to capture the causal relationships between data distribution features and semantics. To further advance the field of semantic communication, it is essential to develop a comprehensive set of semantic-level evaluation metrics.

\section*{Acknowledgment}

We would like to express our sincere gratitude to Chulong Liang, Wei Zhao and Liguang Li from the Algorithm Department of ZTE Corporation for their invaluable support and insights during this research. Their expertise and guidance significantly contributed to the development of this work.

\ifCLASSOPTIONcaptionsoff
  \newpage
\fi

\end{document}